\documentclass[fleqn,useAMS,usenatbib]{mnras}

\usepackage{newtxtext,newtxmath}

\usepackage[T1]{fontenc}
\usepackage{ae,aecompl}

\usepackage{graphicx}	
\usepackage{amsmath}	
\usepackage{amssymb}	

\usepackage{color}
\usepackage{xspace}
\usepackage[below]{placeins}

\newcommand{\msunh}{\mathrm{M}_\odot h^{-1}}
\newcommand{\mpch}{{\rm Mpc}\,h^{-1}}
\graphicspath{{extra_figures/}{figures/}}

\newcommand{\myplot}[1]{\includegraphics[width=0.5\textwidth]{#1}}
\newcommand{\myplottwo}[2]{\myplot{#1}\myplot{#2}}
\newcommand{\mytab}{\begin{table}[htb]}
\newcommand{\myfig}{\begin{figure}[htbp]}
\newcommand{\mybibstyle}{mnras}

\newcommand{\hbtp}{\textsc{hbt+}\xspace}

\newcommand{\fof}{\textsc{FoF}\xspace}
\newcommand{\LCDM}{$\Lambda$CDM\xspace}
\newcommand{\vmax}{V_{\rm max}/V_{\rm vir}}
\newcommand{\onehalf}{{\:\!\!1\!/\:\!\!2}}

\newcommand{\rev}[1]{{#1}}
\newcommand{\revtwo}[1]{{#1}}
\newcommand{\revthree}[1]{{#1}}

\title[halo bias]{The multidimensional dependence of halo bias in the eye of a machine: a tale of halo structure, assembly and environment}
\author[J. Han et al.]{Jiaxin Han,$^{1,2}$\thanks{jiaxin.han@sjtu.edu.cn}
Yin Li,$^{2,3}$
Yipeng Jing,$^{1,4}$
Takahiro Nishimichi,$^2$
Wenting Wang,$^2$
\newauthor{ and Chunyan Jiang$\,^5$}\\
$^1$ Department of Astronomy, Shanghai Jiao Tong University, Shanghai 200240, China\\
$^2$ Kavli IPMU (WPI), UTIAS, The University of Tokyo, Kashiwa, Chiba 277-8583, Japan\\
$^3$ Berkeley Center for Cosmological Physics, Department of Physics, \& Lawrence Berkeley National Laboratory,\\ University of California, Berkeley, CA 94720, USA\\
$^4$ IFSA Collaborative Innovation Center, Shanghai Jiao Tong University, Shanghai 200240, China\\
$^5$ CAS Key Laboratory for Research in Galaxies and Cosmology, Shanghai Astronomical Observatory, Shanghai 200030, China}

\begin{document}
\label{firstpage}
\maketitle


\begin{abstract}
We develop a novel approach in exploring the joint dependence of halo bias on multiple halo properties using Gaussian process regression. Using a \LCDM $N$-body simulation, we carry out a comprehensive study of the joint bias dependence on halo structure, formation history and environment. We show that the bias is a multivariate function of halo properties that falls into three regimes. For massive haloes, halo mass explains the majority of bias variation. For early-forming haloes, bias depends sensitively on the \rev{recent mass accretion history}. For low-mass and late-forming haloes, bias depends more on the structure of a halo such as its shape and spin. Our framework enables us to convincingly prove that $\vmax$ is a lossy proxy of formation time for bias modelling, whereas the mass, spin, shape and formation time variables are non-redundant with respect to each other. Combining mass and formation time \rev{largely} accounts for the mass accretion history dependence of bias. Combining all the internal halo properties fully accounts for the density profile dependence inside haloes, and predicts the clustering variation of individual haloes to a $20\%$ level at $\sim 10\mpch$. When an environmental density is measured outside $1\mpch$ from the halo centre, it outperforms and largely accounts for the bias dependence on the internal halo structure, explaining the bias variation above a level of $30\%$. 
\end{abstract}


%
%
%
%
%

\section{Introduction}
The bias of dark matter haloes describes how haloes with different properties trace the underlying density field, and thus serves as a crucial link between the distribution of haloes and the matter distribution of the universe. It is important for theoretical modelling of the large scale structure, as well as for observational analysis of the distribution of galaxies in the framework of a halo model~\citep[see e.g.][for reviews]{HaloModel, BiasRev}.

How this quantity depends on the mass of haloes is well established in the extended Press-Schechter (EPS) theory by modelling the conditional halo mass function in a given local environment density~\citep{MW96}, or by considering the modulation of the peak counts above a given threshold that is governed by background density (i.e., peak-background split, \citealp{BBKS,CK89}; see also \citealp{Paranjape12,Paranjape13} for some recent extension efforts). 
By construction, these models focus on explaining the bias dependence marginalised over all other quantities except mass.

Beyond the mass dependence, the most well known dependence of bias is the so called ``assembly bias'', that describes the bias dependence on the assembly history of haloes. Parametrising the assembly history with a single formation time parameter, such an assembly bias in low mass haloes ($M<M_\ast$, the collapse mass scale) was first highlighted by \citet{Gao05} using the Millennium simulation~\citep{Millennium}. The strength of this dependence varies with the definition of formation time~\citep{Li08}, which is generally strong at the low mass end but weak~\citep{Jing07} or absent for massive haloes ($M>M_\ast$).

By contrast, significant dependence on other properties of haloes including concentration, spin, shape and dynamical structure has been observed at both the high and low mass end~\citep[e.g.,][]{Wechsler06, Jing07, Bett07, Gao07, Faltenbacher10}. 
Because the structure of a halo is expected to be related to its formation history, the name ``assembly bias'' has been sometimes used to collectively describe the bias dependence on all the remaining parameters besides mass, in spirit of an effort to look for a unified explanation of the various bias dependences. While the connection between halo concentration and mass accretion history has been extensively studied~\citep[e.g.,][]{NFW97,Wechsler02,Zhao03a,Zhao03b,Ludlow13}, for some other halo parameters it is yet to be shown how they can be determined from the halo assembly history. This makes it difficult to prove that assembly history could explain all the bias dependences. In particular, given the weak to no dependence on formation history but much stronger dependence on other structure parameters in cluster haloes, \citet{Mao17} argued that it is inappropriate to collectively call these secondary biases as ``assembly bias''.

There has also been efforts to explain many of the dependences with an environmental variable, because the environment is expected to be important in shaping many of the internal properties of haloes~\citep[e.g.][]{Hahn07b, Hahn07a, Hahn09, Wang11, Lee17}.
For example, \citet{Wang07} quantified the environment of haloes using their expected collapse mass in the initial density field and showed that it can explain the formation time dependence well in low mass haloes. They found that the growth of small haloes in the neighbourhood of massive ones are suppressed by the tidal field from their neighbours, leading to an early formation time and a higher bias as determined by the bias of the massive neighbours. \citet{Salcedo17} also found that the age and concentration dependence of halo bias can be largely expressed as a neighbour bias, while the spin dependence appears to have a different origin. A more straightforward way to quantify the environment is to use the local density measured on a scale significantly larger than the halo scale but much smaller than the scale that defines the bias. According to the peak-background split model, such a density is the quantity that determines the bias of haloes more than halo scale features such as mass or peak height. \citet{Pujol17} showed that such an environment serves as a good proxy to determine the bias of haloes, independent of halo mass. An analytical model for how the bias depends on such an environmental density in addition to halo mass has also been put forward by \citet{Shi18}. 
The morphology of the environment also plays a role in determining the bias, as shown by \citet{Yang17} who measured the bias of haloes in various types of environments including clusters, sheets, filaments and voids~\citep[see also][]{Fisher18}. \citet{Borzyszkowski17, Aseem17} have also studied the environmental dependence of bias focusing on  the effect of an anisotropic tidal field arising from the cosmic web, with an analytical model proposed in \citet{Musso18}.

So far almost all the analysis of the secondary biases take a coarse splitting approach that is optimised mostly for detecting a weak and noisy signal. In this approach, to study the bias dependence on a new variable $x$ besides mass, $M$, one first selects two subsamples of haloes with the highest and lowest $x$ values respectively, and then compares whether and how the average bias as a function of mass (by binning in $M$) differs in the two subsamples. While such an approach allows one to maximise the signal if bias depends monotonically on $x$, it does not provide enough information about the detail of the dependences. Such an approach does not completely disentangle the dependences on $x$ and $M$ either. In case $x$ and $M$ correlates tightly with each other, an apparent signal can be introduced simply due to the finite mass bin size and the correlation between $x$ and $M$, because the high $x$ haloes will preferentially have higher masses than the low $x$ ones in the same mass bin. These limitations make it difficult to analyse the interplay between various bias factors. 
As a result, despite the large number of works that have revealed various bias dependences and the many efforts to find unified explanations of them, it remains largely unclear how these dependences are complimentary or redundant to each other.

To get a clear picture of the interplay between various halo properties in determining halo bias, and to move forward in finding a unified explanation,  we have carried out a systematic analysis of the multivariate dependence of halo bias. In this work, we develop a novel approach that directly computes the joint dependence of bias on multiple variables in an efficient and flexible way. We start by defining a microscopic bias for each halo according to its large scale density profile, and subsequently average the microscopic biases in bins of various multiple parameters to obtain full bias maps. Such an approach also enables us to quantify the sensitivity of the bias to various parameter combinations through a correlation analysis.

In addition, we will apply non-parametric fitting to the bias maps using a popular machine learning technique called Gaussian Process Regression (GPR)~\citep{GPR}. This enables us to construct non-parametric bias estimators in various parameter space, to visualise the multidimensional dependence of bias in interesting projections, as well as to discover parameter redundancies in the bias dependence. Because the bias depends nonlinearly on most halo properties, linear dimensionality reduction methods such as principle component analysis are not applicable for constructing effective parameters. In contrast, our GPR method combined with our sensitivity analysis provides a general non-linear dimensionality reduction framework.

In this work, we focus on extracting and presenting the phenomenologies of the multivariate bias dependences on a few popular and representative parameters describing halo structure, formation history and environment. We leave more physical modelling of these results to future analysis. During the preparation of this work, \citet{Xu17} posted a study on the dependences of bias on a few halo properties that are assumed to be related to halo assembly history. They took a classical approach to measure halo bias from correlation functions, and focused on presenting the two dimensional dependences on mass and another ``assembly variable'' at a time, as well as dependences on two assembly variables at a fixed mass~\citep[see also][for a similar analysis with crude binning]{Lazeyras16}. Compared with their works, our method is more advanced and our analysis is much more comprehensive and systematic. We have not only presented the bias dependences in more than two dimensions on a much larger family of halo properties, but also extracted these dependences as bias estimators and explored their connections and differences.

This paper is organised as follows. We first describe the simulation data including the various halo properties in section~\ref{sec:data}. In section~\ref{sec:method} we introduce the methodology of our analysis. The bias dependence on a single halo property is presented in section~\ref{sec:1d}. For the higher dimensional dependences, we first focus on presenting the joint dependence on structure parameters of haloes in section~\ref{sec:internal}, and move on to analyse further joint dependence on formation time in section~\ref{sec:mah}. In section~\ref{sec:env} we explore the role of the non-parametric density profile on different scales in determining the bias, and discuss the natural definition of an environmental density. We summarise and conclude in section~\ref{sec:summary}.

\begin{figure}
 \myplot{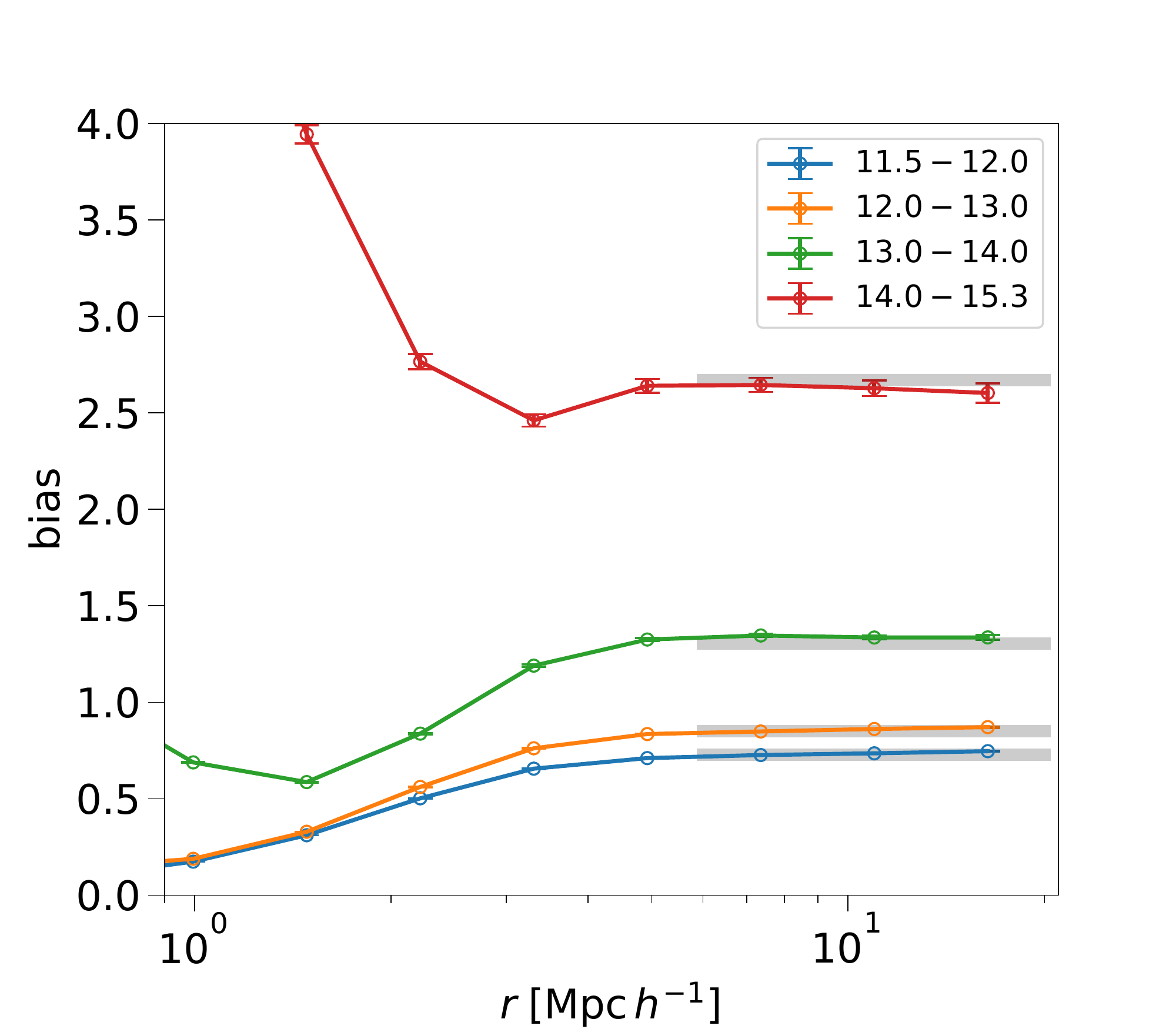}
 \caption{The mass and scale dependence of halo bias, measured from halo-matter cross correlation. Different lines show the halo bias for haloes in different mass ranges, as labelled by $\log (M_{\rm vir}/\msunh)$ in the legend. \rev{The errors shown are errors on the mean bias in each bin, estimated from the scatter in the individual density profile around each halo. For reference, the horizontal thick lines show the expected linear bias value in each bin according to the fitting function of \citet{Pillepich10}.}}\label{fig:mass_scale}
\end{figure}

\section{Data}\label{sec:data}

We use a \LCDM simulation with a boxsize of $600 {\rm Mpc} h^{-1}$ containing $3072^3$ dark matter particles, with cosmological parameters $\Omega_m=0.268$ and $\Omega_\Lambda=0.732$. The simulation started at an initial redshift of $144$ and outputs 100 snapshots uniformly spaced in the logarithm of the scale factor between $z=16.9$ and $z=0$. It is one of the set of CosmicGrowth simulations~\citep{CosmicGrowth} run using a P$^3$M code~\citep{Jing02}. Haloes are identified with the Friends-of-Friends algorithm~\citep{FoF} (\fof) with a standard linking parameter of $0.2$. These haloes are subsequently processed with \hbtp\footnote{\url{https://github.com/Kambrian/HBTplus}}~\citep{HBT,HBTplus} that tracks their evolution throughout the simulation outputs to obtain subhaloes and their evolution histories. In order to reliably resolve the internal structure of a halo, we restrict our analysis to haloes with $M_{\rm vir}>10^{11.5}\msunh$, corresponding to about 500 particles inside the virial radius. This leads to a sample of $2\times 10^6$ haloes with masses up to $3\times 10^{15}\msunh$.

We start from a list of properties describing different aspects of the halo structure, and select a subset of properties that are sensitive bias predictors but not significantly degenerate with each other. The final set of halo properties include halo mass, shape, spin, concentration, and formation history. In addition, we will also study the non-parametric density profile at different scales, including an environmental scale. For the majority of halo properties, we will use the quantity calculated using the smoothly distributed bound particles in the host halo, that is, using particles belonging to the central subhalo found by \hbtp. This is to avoid complications arising from halo substructures. \revtwo{Because the mass contribution from satellite subhaloes to the total halo mass is typically $\sim 10\%$, computing the halo properties using only the central subhalo particles or all the particles around each halo will not lead to any significant difference in most cases. In case they do differ significantly, it reflects the existence of massive subhaloes, which we will not investigate in this study.} For the virial mass and virial radius, however, we compute them using all the particles around the halo, to be consistent with the conventions in the literature.  The centre of each halo is defined as the location of the most bound particle of the central subhalo.

\begin{itemize}
\item $M_{\rm vir}$. The virial mass of the host halo, defined as the spherical mass with a density contrast predicted by the spherical collapse model~\citep[e.g.,][]{BN98}.
\item $V_{\rm max}$. The maximum of the circular velocity function, $V_{\rm circ}(r)=\sqrt{GM(<\!r)\,/\,r}$ of the central subhalo. To separate out the dependence on halo mass, we will mostly use $\vmax$, where $V_{\rm vir}=\sqrt{GM_{\rm vir}/R_{\rm vir}}$ is the circular velocity at the virial radius. This quantity has also been adopted in a few previous studies of halo bias as a proxy for the concentration of haloes, including~\citet{Angulo08,Gao07,Tomomi16}. Note that this ratio can be smaller than 1 given that $V_\textrm{vir}$ is defined from the virial quantities that makes use of all particles inside $R_{\rm vir}$, rather than particles from the central subhalo alone.
\item $e$. The shape parameters defined from the weighted quadrupole tensor, $I_{w,ij}=\sum_p m_p x_{p,i}x_{p,j} / r_{p}^2$, where $m_p$ is the mass of particle $p$, $\vec{x}_{p}$ is the coordinate of particle $p$ relative to the halo centre with $i=1,2,3$ specifying its three components, and $r_p$ is the distance to the halo centre. For the three eigenvalues $\lambda_1\ge\lambda_2\ge\lambda_3$ of the inertial tensor, we define the parameter $e_i=\lambda_i/\sum_{i=1}^3 \lambda_i$. The three eigenvalues specifies the square of the length of the three principle axes of the mass distribution. This weighted inertial tensor has the advantage that it is not dominated by the distribution of distant particles, and thus provides a robust description of the overall directional distribution of particles. We also tried the unweighted (i.e., without $1/r_p^2$ weighting) tensor which gives a similar but weaker correlation with bias. Thus we will only use the weighted tensor hereafter. By construction, only two components out of the three $e_i$ parameters are independent. For real haloes, however, these two components can still be largely correlated. We find that the bias depends weakly on $e_2$ but more on $e_1$ and $e_3$. In addition, both the $e_3$ and $e_2$ dependence of bias can be largely accounted for by the $e_1$ dependence. Thus we only focus on $e_1$ which is denoted simply as $e$.
\item $j$. The spin of the central subhalo following the \citet{Peebles69} definition, $j=\frac{L\sqrt{|E|}}{GM^{5/2}}$, where $L$, $E$ and $M$ are the total angular momentum, energy and mass of the central subhalo. 

\item $a_{M/M_0}$. The scale factor of the universe when the halo mass was a fraction $M/M_0$ of its final value. We sample the mass accretion history (MAH) of each halo with 10 data points, uniformly spaced in the logarithm of halo mass between $0.01 M_0$ and $M_0$, where $M_0$ is the mass of the halo at $z=0$. We record both the halo mass and the scale factor at each sampling point. For an ejected or fly-by halo~\citep[e.g.,][]{Ludlow09}, its \fof and virial masses become ill-defined at the time it was embedded inside a much bigger halo. To avoid such complications, we adopt the bound mass as the default mass definition for the MAH. In particular, we will focus on the scale factor when the halo was half its final mass, $a_\onehalf$, which is obtained by linear interpolation of the MAH using logarithm of both mass and scale factor. In most cases, the interpolation is done using the MAH sampled near $M/M_0=0.4$ and $M/M_0=0.6$.


\item $\delta(r)$. The density profile, $\rho(r)$, of the halo sampled at different comoving radius, expressed in terms of an overdensity $\delta(r)=\rho(r)/\rho_b-1$ where $\rho_b$ is the average matter density of the universe.
For reasons that will become clear later (section~\ref{sec:env}), we will focus on the density $\delta_e=\delta(r_e)$ with $r_e\approx 1\sim2 \mpch$ as a measure of halo environment.

\end{itemize}



\section{Method}\label{sec:method}
\subsection{Microscopic definition of halo bias}
The bias of haloes can be estimated from the halo-matter cross correlation function which is the average density profile around a given population of haloes. To explore the dependence of halo bias on halo properties in a flexible way, we start by defining a microscopic bias of each halo as\footnote{A related halo-to-halo bias in Fourier space has been recently proposed by \citet{Aseem17}.}
\begin{equation}
 \beta(r)=\frac{\delta(r)}{\xi_{mm}(r)},\label{eq:bias_def}
\end{equation} where $\delta(r)$ is the overdensity of matter at radius $r$ around the halo, and $\xi_{mm}(r)$ is the matter-matter correlation function.

$\beta$ is expected to be a very noisy estimate of the underlying bias of the halo, and should only be used in a statistical way.
\rev{For a specific population of haloes, our estimator simply reduces to the usual
linear bias once ensemble averaged $b(r) = \langle \beta(r) \rangle =
\xi_{hm}(r) / \xi_{mm}(r)$. }

In general, $\beta$ depends on the specific halo population characterised by their
properties. Let $\mathbb{A}$ be the full set of inherent halo properties determining their bias.
Thus $\beta$ can be a function of $\mathbb{A}$ plus a random component that is
independent of $\mathbb{A}$, i.e.\ $\beta = f(\mathbb{A}) + \epsilon$.
The average bias as a function of a subset of properties, $\mathbb{B}\subset\mathbb{A}$, is obtained by
\begin{align}
b(\mathbb{B})&=\int \beta(\mathbb{A}) \mathrm{d}P(\mathbb{A}-\mathbb{B}|\mathbb{B})\nonumber\\
&=\int b(\mathbb{A}) \mathrm{d}P(\mathbb{A}-\mathbb{B}|\mathbb{B}),\label{eq:marginal}
\end{align}
which is the full bias function marginalised over the complementary set $\mathbb{A}-\mathbb{B}$. As an example, suppose the full bias depends on all the halo properties listed in section~\ref{sec:data}. Then the mass dependence of bias can be written as $b(M_{\rm vir})=\int b(M_{\rm vir}, \vmax, e, ...) \mathrm{d}P(\vmax, e, ...)$, where the marginalisation is done over distribution of all the halo properties except mass.


Fig.~\ref{fig:mass_scale} shows the average bias of haloes binned in mass and radius. Overall, haloes are more biased above the characteristic mass scale $M_\ast\approx10^{12.5}\msunh$ and anti-biased below $M_\ast$. For the radial range of $5-20 \mpch$, the bias is consistent with being a constant. \rev{To see that this bias estimate is also the linear bias factor usually measured on even larger scales, we  compare our result with the fitting function of \citet{Pillepich10}. \citet{Pillepich10} measured the bias of \fof haloes over the scale $0.01<k<0.05 h/{\rm Mpc}$ from the halo-matter correlation function, and fitted this linear bias as a function of \fof halo mass. To compare with their measurement, we compute the bias of each halo in our sample using their fitting function, and show the average of the predicted bias in each bin as a horizontal thick line. The predicted linear bias values agree well with our measurements for $r>5\mpch$. For our measurements, the signal to noise decreases as one goes to a larger scale. Note that on large scales, correlation among the individual density profile becomes increasingly important, and the errors are subsequently underestimated.} As a compromise between signal to noise and the scale invariance,  \rev{we will measure the microscopic bias in the radial bin of $6-9\mpch$ following Equation~\eqref{eq:bias_def} hereafter}, and drop the explicit radial dependence in $\beta(r)$. Adopting a larger radial scale does not affect any of our conclusions, except for the exact values of the sensitivities that we discuss below.

\subsection{Sensitivity analysis}\label{sec:sensitivity}
We use the correlation coefficient to explore the response of halo bias to variations in halo properties. The Pearson correlation coefficient between two variables $x$ and $y$ is defined as
\begin{equation}\label{eq:cov}
 \gamma_{x,y}=\frac{\langle (x-\bar{x})(y-\bar{y})\rangle}{\sigma_x\sigma_y},
\end{equation} where $\bar{x}$ and $\sigma_x$ are the mean and standard deviation of $x$. 

In the context of halo bias, the correlation coefficient between $\beta$ and a generic halo property $x$ is
\begin{align}
 \gamma_{x, \beta}&=\frac{\langle (x-\bar{x}) (\beta-\bar{\beta})\rangle}{\sigma_x\sigma_\beta}\nonumber\\
 &=\frac{\langle \delta_x \beta \rangle}{\sigma_x\sigma_\beta}\nonumber\\
 &=\frac{\langle \delta_x b(x)\rangle}{\sigma_x\sigma_\beta},\label{eq:corr_bias}
\end{align} where we have used $b(x)=\langle\beta|x\rangle$ and $\delta_x=x-\bar{x}$.

In the case that $b(x)=kx/\sigma_x+b_0$, then $\gamma=k/\sigma_\beta$, so the correlation coefficient directly measures the steepness of the (normalised) bias as a function of the normalised halo property $x/\sigma_x$. It can also be understood as the fractional bias variation that can be explained by $x$. For one standard deviation in $x$, the induced variation in bias is $\Delta b(\Delta x=\sigma_x)=k$, so that $\gamma=\Delta b(\Delta x=\sigma_x)/\sigma_\beta$. 


More generally, assuming $\beta=\sum k_i x_i / \sigma_{x_i} +\epsilon$, where $\{x_i\}$ are independent variables describing the haloes, it is easy to see that the correlation operator is the projection operator in the dependency space, $\gamma_{x_i, \beta}=k_i/\sigma_\beta$, and measures the fraction of the variation in $\beta$ that can be predicted from $x_i$. Later in this paper, we will continue to use $x$ to represent a generic halo property.



The error of the correlation coefficient can be estimated as\citep[e.g.,][]{Bowley28} $e_c=(1-\gamma^2)/\sqrt{1-N}$ where $N$ is the sample size. Alternatively, one can estimate the error using the bootstrap method. Another way is to compute it from the distribution of correlation coefficients of shuffled datasets in which $x$ is permuted randomly. We have checked that all three methods give consistent results, and thus we will simply use the analytical estimate.


\begin{figure*}
 \myplottwo{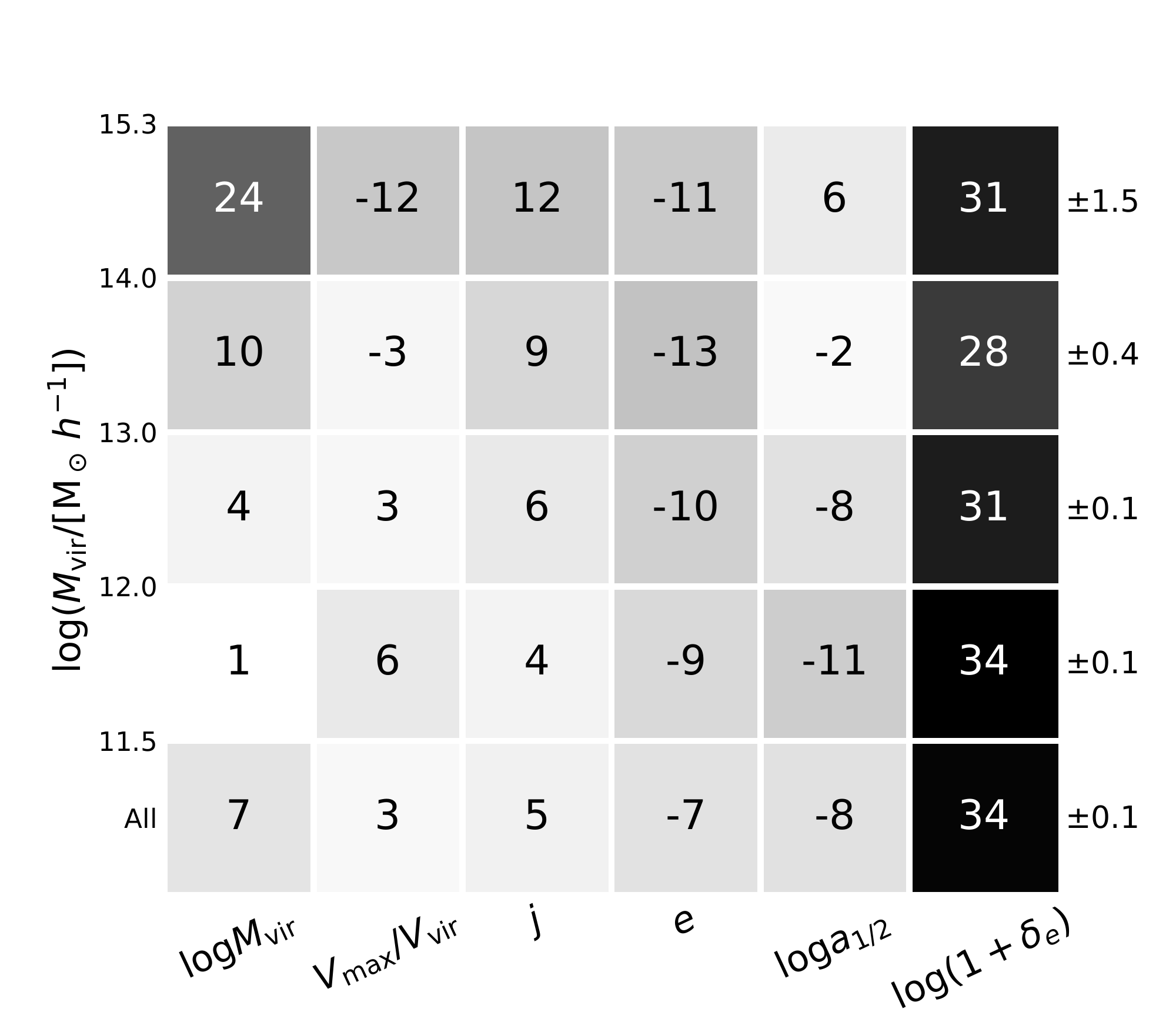}{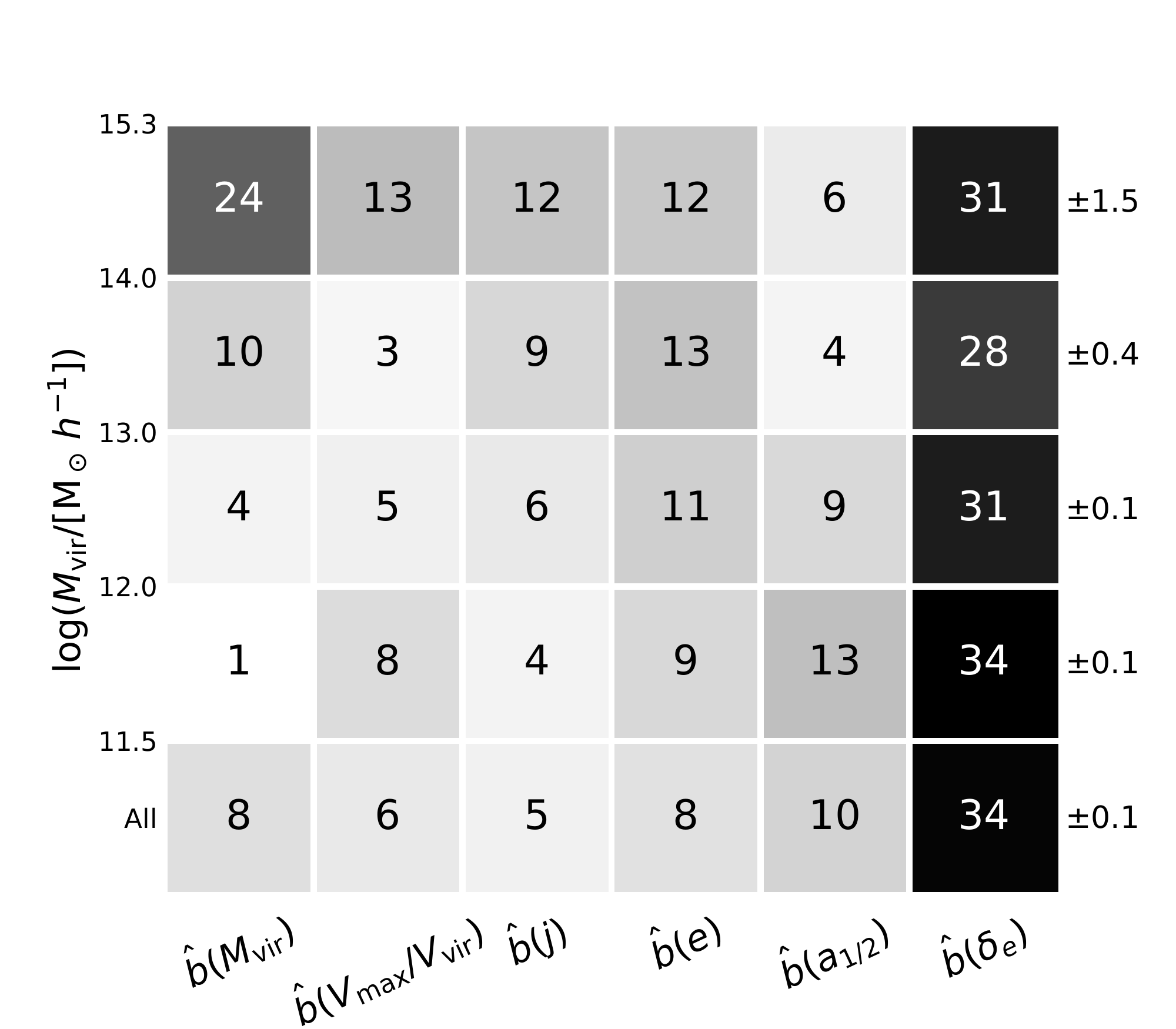}
 \caption{\textit{Left}: sensitivity of halo bias to various halo properties listed in section~\ref{sec:data}. The number in each cell shows the correlation coefficient, $\gamma$ (Eq.~\ref{eq:corr_bias}), expressed in units of percent, which quantifies the fraction of the variation in clustering that can be explained by the variation of the given halo property. Each column shows the sensitivity to one halo property, while each row shows the sensitivity measured in one halo mass range, with the last row (`All') measured using the full sample. The numbers on the right show the uncertainty in the correlation coefficient for each row in the same unit. \textit{Right}: same as the left except that each variable has been replaced by its bias estimator when computing the correlation coefficient. }\label{fig:sens_all}
\end{figure*}

The absolute value of $\gamma$ depends on $\sigma_\beta$ which can be affected by the sampling noise in $\beta$ (as determined by, e.g., the radial binning and the numerical resolution of the simulation), in addition to the radial scale of the bias measurement. However, once the radial scale and sampling are fixed, the relative amplitude between different $\gamma$ reflects the relative sensitivity of bias to different variables. For our sample, $\sigma_\beta\simeq 2$ and is nearly independent of selections. We have also checked that the Poisson noise in $\beta$ associated with our sampling 
is only at the level of $0.1\%$ thus negligible. As a result, our correlation estimate should reflect the intrinsic correlation to bias at the scale of $6-9\mpch$. If one is interested in comparing our result to that at a different (typically larger) scale, it is straightforward to convert $\gamma$ to the amount of predictable bias variation over the typical variation of $x$ as $\Delta\beta_x=\gamma_{\beta,x}\sigma_\beta$, which is then independent of sampling effects and radial binning. Except for the exact values in $\gamma$, none of our conclusions are sensitive to the choice of the radial scale of the bias measurement.

\subsection{Gaussian Process Regression}
By averaging $\beta$ in bins of one or multiple halo properties $(x, y, ...)$, we can obtain pixelised bias functions or bias maps in the halo property space, $b(x, y, ...)=\langle \beta|x,y,... \rangle$. We will use Gaussian Process Regression (GPR) as implemented in the \textsc{Python} package \textsc{scikit-learn}~\citep{scikit-learn}\footnote{\url{http://scikit-learn.org/stable/modules/gaussian_process.html}} to obtain smooth interpolation of the pixelised bias maps and use them as bias estimators for further analysis. We will use $\hat{b}(x,y,...)$ to denote these GPR interpolation functions, which are expected to be accurate representations of the underlying true bias functions $b(x,y,...)$ with negligible deviations.

GPR is a method that fits a Gaussian Process (GP) to data points. The merit of this technique is that it allows for fitting arbitrary data points in multiple dimensions without assuming any functional form. GPR can be used in a Bayesian way to also derive statistical bounds on the fits. In this work, we use GPR simply as a flexible non-parametric smooth fit to the multidimensional data points. Even though we use GPR extensively throughout this paper, a thorough understanding of this technique is not essential for understanding the results of this paper, as long as one recognises GPR to be a non-parametric interpolation method. We briefly explain how it works below, and provide the mathematical details for interested readers in Appendix~\ref{app:GPR}.

In our context, a GPR aims to reconstruct a Gaussian random field (or GP) in the space of halo properties, subject to the constraint that the field values are observed to be the bias value in each halo property bin. By further specifying a correlation function (or kernel) of the Gaussian random field with some free hyper parameters, one can write down the likelihood of observing the bias map given any hyper parameter values, and derive posterior distribution of the Gaussian random field that is consistent with the observed bias map and its errors. We adopt a kernel function that yields first-order differentiable Gaussian random fields. The average of the posterior Gaussian random field then serves as a smooth function that interpolates through the bias map. 

\begin{figure*}
 \includegraphics[width=\textwidth]{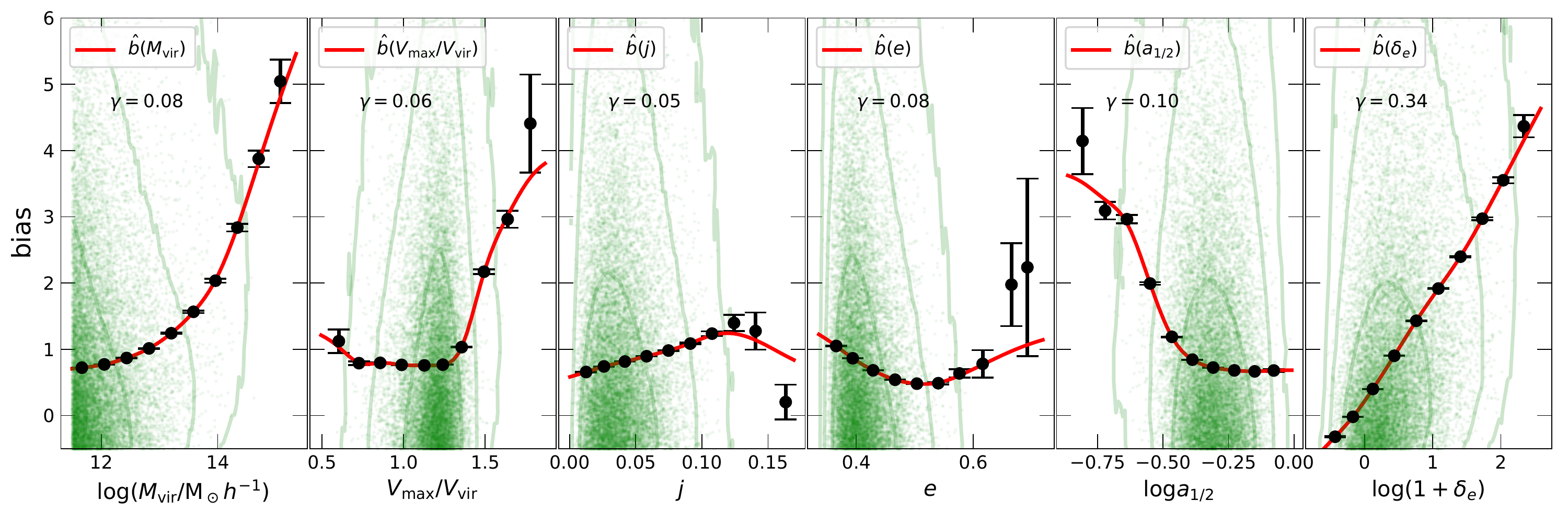}
\caption{The dependence of bias on a single variable for haloes with $M_{\rm vir}>10^{11.5}\msunh$. The light green dots show the individual biases, $\beta$, for one percent of randomly selected haloes. The light green contours are constant-density lines of the halo distribution in each plane enclosing $68.3\%$, $95.5\%$ and $99.7\%$ of the haloes. The data points with error bars are measurements of the mean bias function $b=<\beta>$ in each halo property bin, while the curves are GPR fits to the binned measurements. The number on each plot shows the correlation coefficient between $\beta$ and the GPR prediction (identical to the last row in the right panel of Fig.~\ref{fig:sens_all}).}\label{fig:GPR1d}
\end{figure*}

\section{Dependence on a single halo parameter}\label{sec:1d}

In the left panel of Fig.~\ref{fig:sens_all}, we show the sensitivity of halo bias to different halo properties listed in Section~\ref{sec:data}. The sensitivities refer to the correlation coefficient defined in Equation~\ref{eq:corr_bias}. Because halo mass is often a selection variable for halo samples in both simulations and observations, we also divide the sample into four mass bins and measure the sensitivities in each bin. This figure summarises most of the phenomena that will be explored in detail in the following sections.

It is well known that bias depends weakly on mass at the low mass end, and increases steeply at the high mass end. This is reflected in Fig.~\ref{fig:sens_all} as an increasing sensitivity to halo mass, from zero dependence at the low mass end to $\sim20$ percent explained bias for cluster haloes. The sensitivity to $\vmax$ and formation time $a_\onehalf$, which are well correlated as shown later (see section~\ref{sec:mah}), also evolve monotonically with mass and reverse signs in cluster haloes, consistent with previous findings using percentile split measurements~\citep[e.g.,][]{Gao07,Jing07,Angulo08}. On the other hand, the sensitivities to structure parameters $e$ and $j$ are relatively stable across different mass ranges. In particular, in cluster haloes, the sensitivity to formation time is much weaker compared to the sensitivity to $e$ and $j$, in line with \citet{Mao17} who found a lack of dependence on assembly history in cluster haloes.


Out of all the properties studied in Fig.~\ref{fig:sens_all}, we find that the environment $\delta_e$ is the most sensitive predictor of halo bias, the sensitivity of which is also the most stable across different mass ranges.

We have shown that the correlation coefficient is equivalent to the slope of the bias dependence $b(x)$ if it is linear in $x$. For a strongly nonlinear $b(x)$, however, $\gamma_{\beta, x}$ is not optimal in revealing its sensitivity to $x$. In this case, it is desirable to apply a functional transform to $x$ that leads to a linear dependence of bias on the transformed $x$, and compute the correlation between $\beta$ and the transformed $x$. A bias model as a function of $x$ would naturally provide such a nonlinear transformation. To this end, we first extract the functional form of the bias dependence $\hat{b}(x)$ from the data, and then compute the correlation coefficient $\gamma_{\hat{b}(x), \beta}$. Here $\hat{b}(x)$ is obtained from a GPR fit to the average $\beta$ in bins of $x$. Because $\beta=\hat{b}+\epsilon$ by construction, it is straightforward to show that $\gamma_{\hat{b}(x), \beta}=\sigma_{\hat{b}(x)}/\sigma_{\beta}$, which again measures the amount of modelled bias variation, $\sigma_{\hat{b}(x)}$, as a fraction of the total bias variation, $\sigma_{\beta}$. However, unlike in the case of $\gamma_{x, \beta}$, this interpretation of $\gamma_{\hat{b}(x), \beta}$ no longer requires $\hat{b}(x)$ to be linear.

As shown in Fig.~\ref{fig:GPR1d}, the bias depends non-linearly on most variables except for the environment, and the GPR fits the dependences satisfactorily. \rev{Note that we account for the uncertainty of each data point during GPR fitting. Artificially decreasing the uncertainty of the data points will lead to a better fit, at the expense of potentially overfitting the data.} The correlation coefficient between $\beta$ and the $\hat{b}(x)$ prediction for each halo are shown in the right panel of Fig.~\ref{fig:sens_all}. For each mass range, we fit a new estimator using only haloes in the current mass range, to be consistent with the left panel. The magnitude of the correlation coefficient after the transformation is always larger or equal. The difference is negligible in most cases and does not affect our previous conclusions. Note that the correlation is always positive in $\hat{b}(x)$ space, because the sign of correlation is absorbed into the slope of the $\hat{b}(x)$ function. For the environment $\delta_e$, the bias model $\hat{b}(\log(1+\delta_e))$ is close to linear, so that it makes little difference whether the $\hat{b}(x)$ transformation is applied or not. Similar arguments apply when the mass dependence is studied in each mass range.

The dynamic range of bias probed by mass is one of the largest in Fig.~\ref{fig:GPR1d}. One may wonder why the overall correlation coefficient with mass is not much larger than the others. This is because the mass distribution of haloes is dominated by low mass ones, so that the high mass haloes that show the highest bias do not contribute very much to the correlation. In Fig.~\ref{fig:sens_all}, the correlations estimated in different mass bins help to clarify the situation. For the majority of haloes with lower masses, the variation in $\beta$ is not better explained by the variation in mass than other variables such as $e$, which is also evident from the scatter plot (light green points and contours) in Fig.~\ref{fig:GPR1d}. 

The same is true for the $\vmax$ and $a_\onehalf$ dependences. In addition, the dependences on these two variables also separate into two regimes. The dependences are weak or absent for $\vmax\lesssim1.3$ and $\log a_\onehalf>-0.4$, but strong at high $\vmax$ as well as low $\log a_\onehalf$. \rev{Early-forming haloes are more biased, in qualitative agreement with previous findings~\citep[e.g.][]{Gao05}.}

As noted in Equation.~\ref{eq:marginal}, the various single parameter dependences can be understood as different projections of the same multivariate bias function. A complete picture of how the various bias dependences are related to each other can only be obtained by studying the joint dependences of bias on multiple variables. In the following sections, we will explore the joint dependence of halo bias on internal halo properties, on formation time and on environment step by step.

\begin{figure*}
 \includegraphics[width=0.7\textwidth]{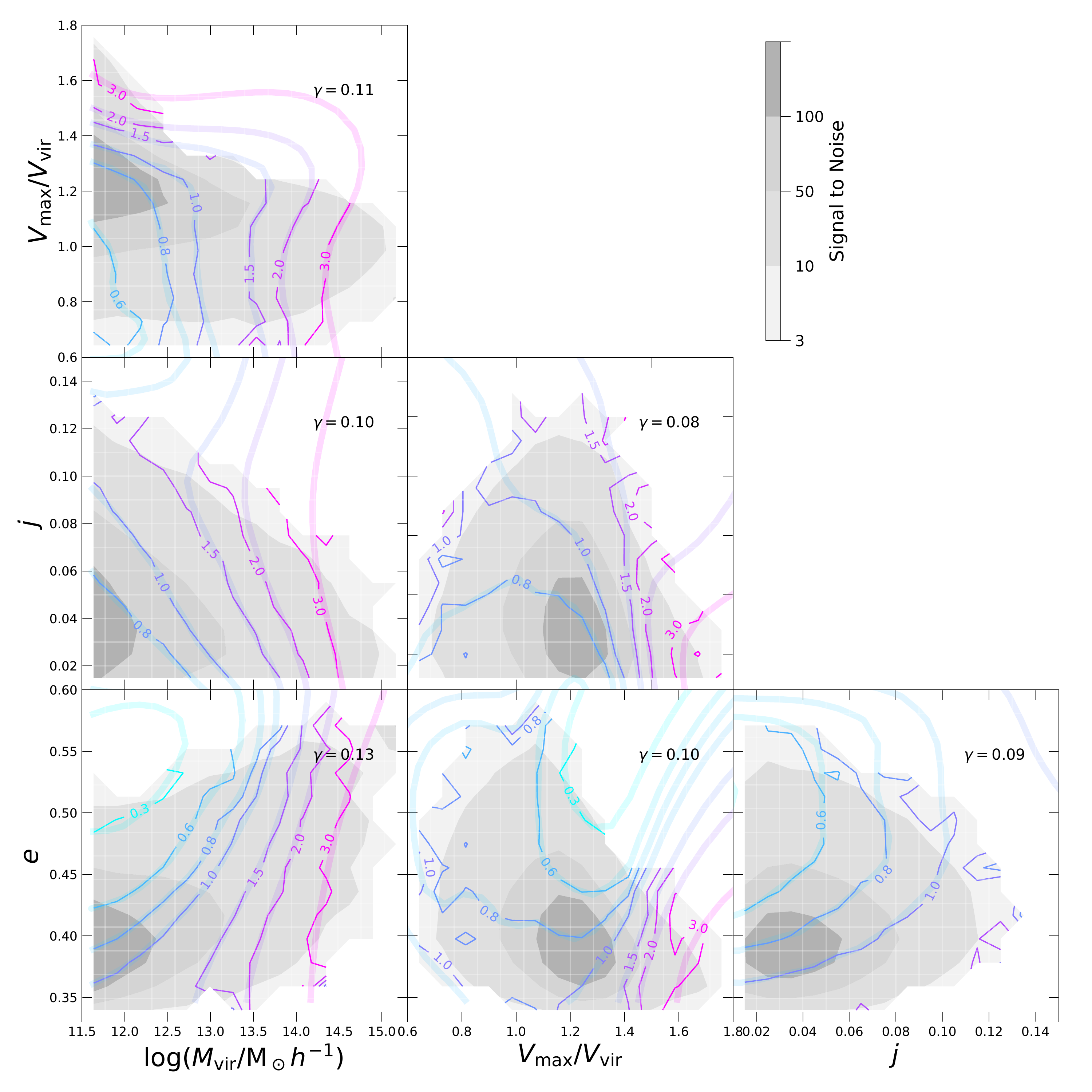}
 \caption{The two dimensional dependence of bias on internal properties of haloes listed in Section~\ref{sec:data}. The thin coloured contours show the bias levels of haloes, while the thick contours show GPR fits, $\hat{b}(x,y)$, to each bias map, where $(x,y)$ refers to the two axes of each map. The grey filled contours show the signal to noise level of the bias estimates at each location, which is mostly determined by the number of haloes inside each bin. The white grids show the binning used in computing the bias map. The number $\gamma$ on each map shows the correlation coefficient between the GPR fit and the microscopic bias $\beta$, \rev{and quantifies the fraction of $\beta$ variation that is modellable by the $\hat{b}(x,y)$ model.}. Only bins with a signal to noise ratio above 3 are displayed, while the GPR fits are shown over the entire map. The apparent discrepancy at the high $j$ end in the $(M_{\rm vir}, j)$ plane is caused by accommodating the lower signal to noise pixels that are not shown.}\label{fig:internal}
\end{figure*}

\section{Joint dependence on halo internal properties}\label{sec:internal}
\subsection{The two dimensional dependence}\label{sec:2d}
With $\beta$ calculated for each halo, it is straight-forward to obtain the joint dependence of bias on any halo parameters by averaging $\beta$ inside multidimensional bins of those parameters. In Fig.~\ref{fig:internal} we show the two-dimensional dependence of bias on the internal properties of haloes. Overall, none of the variables in Fig.~\ref{fig:internal} is completely redundant with respect to the other of the pair, as evidenced by the non-trivial shape of the bias contours.

Combining halo mass and structure parameters, the bias dependence broadly separates into two regimes. At the high mass end ($M\gtrsim 10^{13}\msunh$), the bias depends strongly on halo mass. For low mass haloes, however, the dependences on structure parameters become more prominent, consistent with the picture in Fig.~\ref{fig:sens_all}. 

The joint dependences depicted in Fig.~\ref{fig:internal} also help to gain insights into the one dimensional bias dependences in Fig.~\ref{fig:GPR1d}. In the $(M_{\rm vir}, \vmax)$ space, the bias depends mostly on $\vmax$ for high $\vmax$ haloes, leading to the steep rise in the $b(\vmax)$ function at the high $\vmax$ end in Fig.~\ref{fig:GPR1d}. For low $\vmax$ haloes, bias depends mostly on mass rather than $\vmax$. Because $\vmax$ is hardly correlated with mass for these haloes, this leads to a flat $\vmax$ dependence. In the $(M_{\rm vir}, j)$ space, bias depends positively on both mass and spin everywhere. At the low spin end, mass and spin are uncorrelated, so that the net marginalised spin dependence remains positive. At the high spin end, however, there is a shortage of high mass haloes, so that the bias of the high spin haloes are primarily determined by the low mass ones, leading to a decrease in $b(j)$ at the high spin end. In the $(M_{\rm vir}, e)$ space, bias depends positively on mass but negatively on $e$. At the same time, the high $e$ haloes are mostly massive ones, which leads to a rise in $b(e)$ at the high $e$ end. Similar arguments can be applied to go from other joint distributions in Fig.~\ref{fig:internal} to the corresponding marginalised distributions in Fig.~\ref{fig:GPR1d}. Such exercises demonstrate that the joint dependences are more intrinsic, while the marginalised dependences result from both the joint dependences and the distribution of the halo sample in the parameter space (see Equation~\ref{eq:marginal}).

\begin{figure*}
 \includegraphics[width=\textwidth]{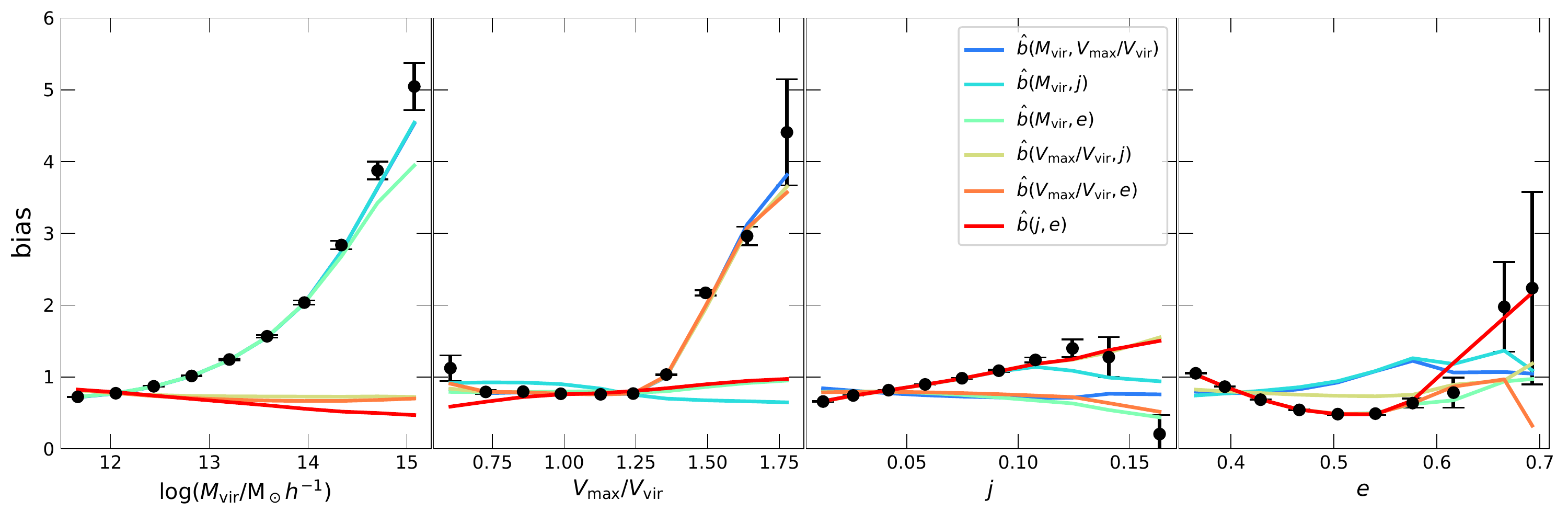}
 \caption{The performance of bivariate estimators in predicting the marginalised bias dependence. Same as in Fig.~\ref{fig:GPR1d}, the data points with errorbars are the measured bias dependence on a single variable for haloes with $M>10^{11.5}\msunh$. Lines with different colours are the predictions from various estimators that makes use of two halo properties.}\label{fig:GPR2d}
\end{figure*}

Similar to the exercise in Fig.~\ref{fig:GPR1d}, we have also performed GPR fits to the two dimensional bias maps, to obtain bias estimators combining two halo variables. 
Although only bins with a signal to noise above 3 are shown in Fig.~\ref{fig:internal}, for the fitting we use all bins with a signal to noise above 1 for better accuracy near boundaries of the parameter distributions. Again these GPR fits well represent the two dimensional dependence of bias. The apparent discrepancy at the high $j$ end in the $(M_{\rm vir}, j)$ plane is caused by accommodating the lower signal to noise pixels that are not shown. The correlation coefficients between $\beta$ and these bivariate estimators are labelled in the figure. Compared with the one dimensional estimators in Fig.~\ref{fig:GPR1d}, adding a new variable always increases the correlation, implying increased performance in predicting the bias. The $\hat{b}(M_{\rm vir}, e)$ estimator has the highest sensitivity, although with only a minor gain compared to the other bivariate estimators. 

Each bivariate estimator models the joint dependence on two variables. If such a model correctly captures the dependence on both variables, it should be able to reproduce the marginalised dependences on each of the two. Moreover, if the bias dependence on a third variable can also be explained by the dependence on the two variables modelled, one would also expect the bivariate estimator to reproduce the marginalised dependence on the third variable. We present such tests in Fig.~\ref{fig:GPR2d}.
For each estimator, we first apply it to each halo to predict a bias value, and then average the predictions for all the haloes in bins of a given halo property to obtain marginalised predictions in that dimension.
Not surprisingly, each bivariate estimator successfully reproduces the marginalised dependence on its own variables. The slight disagreements in bins with large error bars are due to the small number of haloes available in these bins, so that the two dimensional sampling used in constructing the estimator is too noisy. 
On the other hand, all the estimators struggle to reproduce the bias dependence on variables that are absent in its construction. The predicted bias of the left-out variable is mostly flat, that is, almost no dependence can be predicted if the variable is not modelled explicitly. This means the bias dependences on the four variables are largely \emph{independent} of each other. We will further demonstrate this independence in higher dimensions below.

\begin{figure*}
 \includegraphics[width=0.5\textwidth]{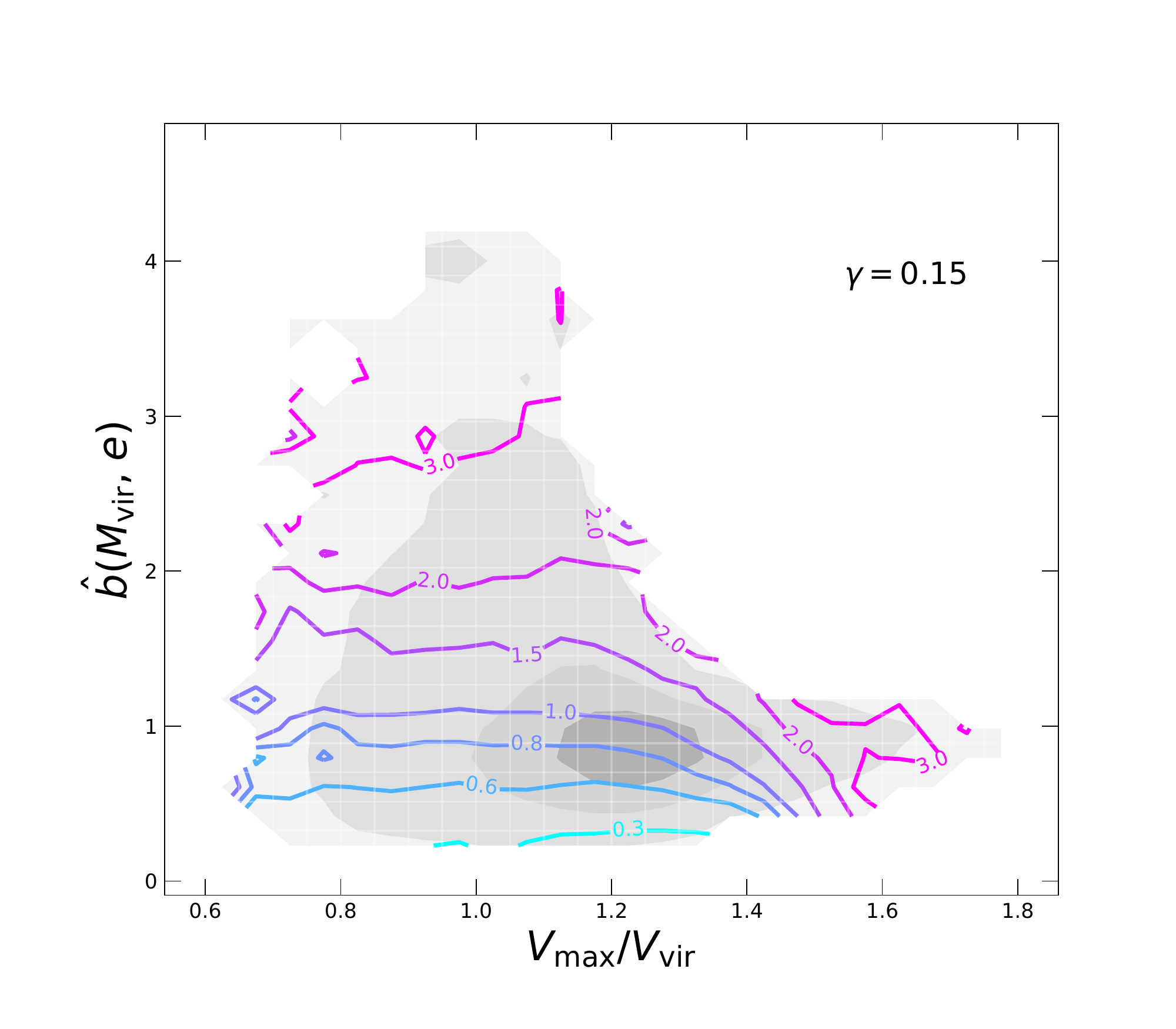}\includegraphics[width=0.5\textwidth]{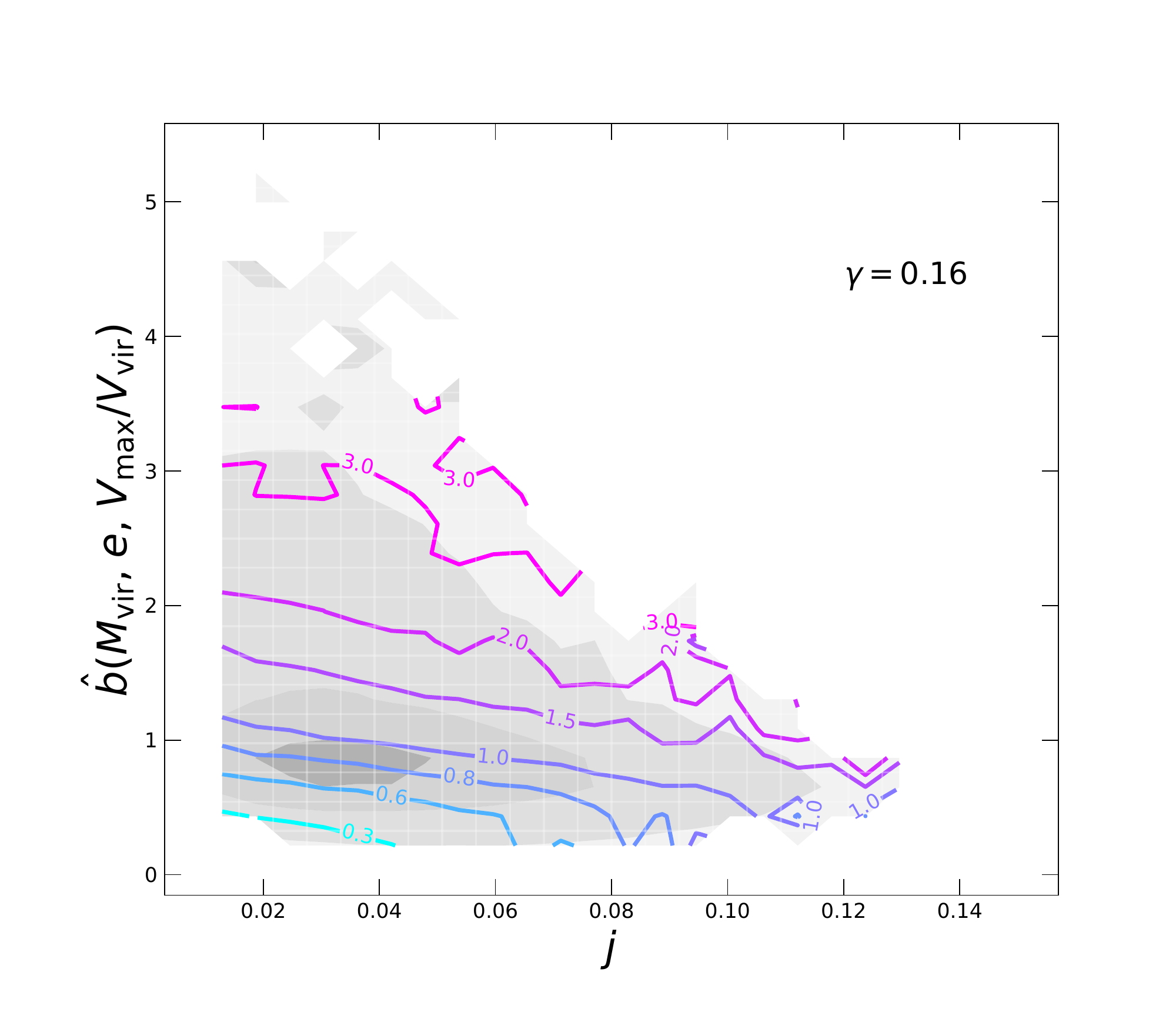}
 \caption{\textit{Left}: the joint dependence of bias on a bivariate estimator, $\hat{b}(M_{\rm vir},\;e)$, and $\vmax$. \textit{Right}: the joint dependence of bias on the three dimensional estimator constructed from the left panel, $\hat{b}(M_{\rm vir},\;e,\;\vmax)$, and the left-out halo property $j$. In each panel, the $\gamma$ value shows the correlation coefficient between the GPR estimator constructed from this map and the bias field $\beta$, \rev{and quantifies the fraction of $\beta$ variation that is modellable by the bias estimator}. For clarity, we have not shown the GPR contours.}\label{fig:map4d}
\end{figure*}

\subsection{Beyond two dimensional dependence: the non-redundancy of structure parameters}
One can straightforwardly generalise the 2D method to construct bias maps and GPR estimators into higher dimensions.
However, the statistical significance quickly starts to suffer from the curse of dimensionality with more variables added.
The sampling of the bias function in higher dimensions becomes more and more challenging. This is because the number of haloes located near the boundaries of the parameter space increase exponentially with its dimension.  At the same time, for a fixed bin width, the number of haloes in each bin also decreases exponentially with dimension.
In addition the maps also become difficult to visualise beyond two dimensions.

To sidestep these problems, we first compute 2D maps in the parameter space of the GPR'ed $\hat b(x, y)$ and a third property $z$, and then apply GPR again to these maps to construct three dimensional estimators of the recursive form $\hat{b}(\hat{b}(x,y),z)$. We show one example of this approach in the left panel of Fig.~\ref{fig:map4d}. This time we treat $\hat b(M_{\rm vir}, e)$ constructed in section~\ref{sec:2d} as a derived property of each halo, and compute the bias map of it together with $\vmax$. It visualises the residual dependence of bias on $\vmax$ after accounting for the joint $(M_{\rm vir}, e)$ dependence. In the low $\vmax$ end, there is almost no dependence on $\vmax$, in contrast to the complex dependence in the $(e, \vmax)$ space in Fig.~\ref{fig:internal}. This means the apparent $\vmax$ dependence there at fixed $e$ is introduced through the mass dependence. Once we account for the joint $(M_{\rm vir}, e)$ dependence, the remaining $\vmax$ dependence is eliminated at the low $\vmax$ end. This further demonstrates the power of such multidimensional analysis in disentangling the complex dependences on multiple variables.

By applying GPR to this map we can construct an estimator of the form $\hat{b}(\hat{b}(M_{\rm vir}, e), \vmax)$. The corresponding correlation coefficient between this new estimator and $\beta$ is also shown. Its increment compared to $\gamma_{\hat b(M_{\rm vir}, e), \beta}$ tells the additional bias variation explained by $\vmax$. We have constructed such estimators for all triplet combinations of the structure variables and show the results in Appendix~\ref{app:3d}. Out of all the three dimensional estimators, the most sensitive one is $\hat{b}(\hat{b}(M_{\rm vir}, e), \vmax)$ shown in Fig.~\ref{fig:map4d}, which explains the bias variation to 15\%.

\begin{figure*}
 \includegraphics[width=\textwidth]{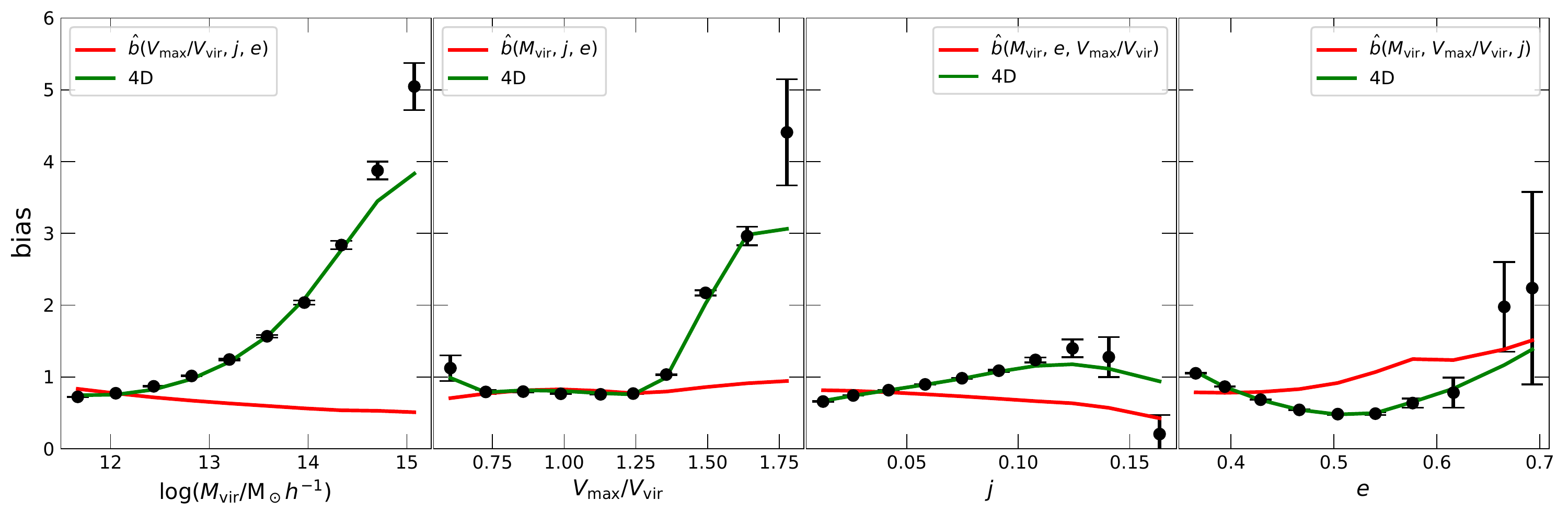}
 \caption{The performance of three and four dimensional estimators in predicting the marginalised bias dependence. For clarity, in each panel we only show one version of the 3D estimator that does not model the $x$-axis variable explicitly. The other permuted 3D variants are almost indistinguishable. The 4D estimator is the same as the one constructed in Fig.~\ref{fig:map4d}.}\label{fig:GPR4d}
\end{figure*}

Note that $\hat{b}(\hat{b}(M_{\rm vir},\, e),\, \vmax)$ is not equivalent to $\hat{b}(\hat{b}(M_{\rm vir},\,\vmax),\, e)$ in general, as they fit different projections of the three dimensional bias manifold. So these nested estimators are only pseudo three dimensional estimators. However, we find that different variants of the pseudo 3D estimators yield very similar $\gamma$ values. As a result, it is reasonable to expect that the full 3D estimators would not improve significantly over the pseudo ones.

Following this procedure we can move on to construct pseudo 4D estimators combining all the internal halo properties. Fig.~\ref{fig:map4d} shows such an effort. With a fourth variable included, the correlation coefficient increases slightly to 16\%. Despite this, this 4D estimator now successfully predicts the marginalised dependence of bias on any of the four halo properties in Fig.~\ref{fig:GPR4d}, in contrast to the 3D estimators which all fail to reproduce the marginalised dependence on the left-out halo property. This demonstrates that \emph{all four halo properties investigated here contain different information about the bias and are non-redundant} with respect to the other three. We have checked that combining the four variables in different orders produces very similar results. For simplicity, from now on we will use $\hat{b}(x,y,z,...)$ to denote the nested estimator $\hat{b}(\hat{b}(\hat{b}(x,y),z),...)$ with three or more variables.


\section{Mass accretion history dependence}\label{sec:mah}
Now we focus on the dependence of bias on the mass accretion history (MAH) of each halo. \revtwo{It is well-known that the MAH of haloes largely follows a universal form~\citep[e.g.]{Zhao03a,Wechsler02}, so that one or two formation time parameters can largely describe the MAH to leading orders.} We start with a parametrization of the MAH with a single formation time parameter, $a_\onehalf$. Later in this section we further investigate whether there are additional bias information in the MAH beyond this formation time.

\begin{figure*}
 \includegraphics[width=\textwidth]{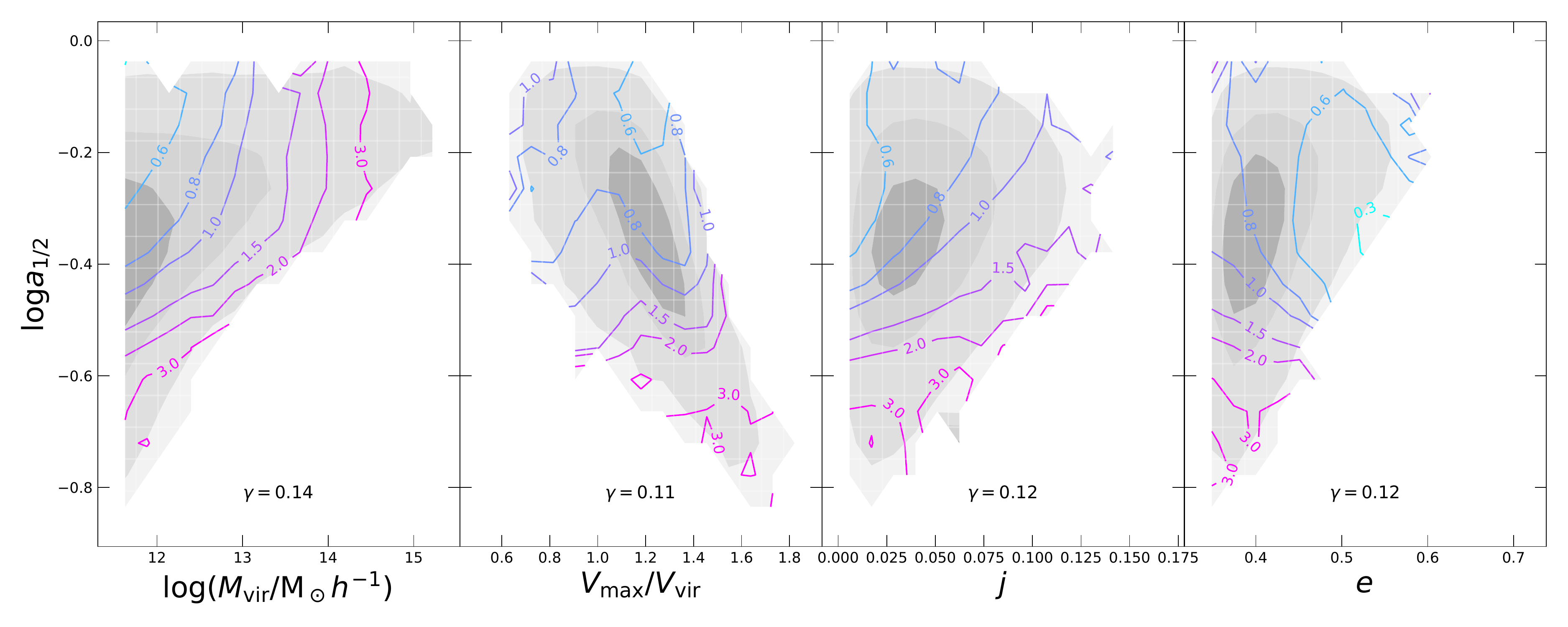}
 \caption{The joint dependence of bias on halo formation time and halo internal properties. Same as in Fig.~\ref{fig:internal}, the grey filled contours show the signal to noise of the bias estimate in each pixel. The thin colour contours show the bias level. The $\gamma$ value show the correlation coefficient between $\beta$ and the GPR fit, \rev{$\hat{b}(x,y)$, to each map, where $(x,y)$ refers to the two axes of the map. It quantifies the fraction of $\beta$ variation that is modellable by $\hat{b}(x,y)$}. For clarity, we have not shown the GPR contours.}\label{fig:bias_mah}
\end{figure*}

\subsection{Formation time dependence}
In Fig.~\ref{fig:bias_mah} we show the joint dependence of bias on halo formation time and other internal halo properties. Similar to the findings in sections~\ref{sec:1d} and \ref{sec:internal}, the joint dependence on mass and formation time broadly separates into two regimes in mass: for massive haloes the bias depends mostly on mass, while for low mass haloes there is a significant dependence on formation time.

Analogously, the formation time dependence also broadly separates into two regimes: for early-forming haloes the bias depends mostly on formation time, while for late-forming haloes the dependence on a second parameter is more significant. Early forming haloes are less massive, close to spherical and slowly rotating. These haloes are highly biased than late-forming haloes with similar internal properties, in sharp contrast to the overall low bias for low mass haloes expected from the mass dependence alone.

\subsubsection{The $\vmax$ -- $a_\onehalf$ connection}
As shown in the second panel of Fig.~\ref{fig:bias_mah}, there is a significant correlation between formation time and $\vmax$ \rev{according to the grey contours.} \rev{These $S/N$ (grey filled) contours in the bias maps are primarily determined by the number of haloes in each pixel, and hence also describes the distribution of haloes in each map, with higher $S/N$ regions populated by more haloes.} Note that $\vmax$ can be interpreted as the shape of the density profile: isothermal haloes have flat rotation curves with $\vmax=1$, while a larger $\vmax$ correspond to a steeper outer profile. The bias dependence on $\vmax$ largely finds explanation in the formation time dependence \rev{given this correlation and the mostly horizontal bias contours (coloured lines)}. Haloes with higher $\vmax$ are formed earlier and more biased.

\begin{figure*}
 \includegraphics[width=0.4\textwidth]{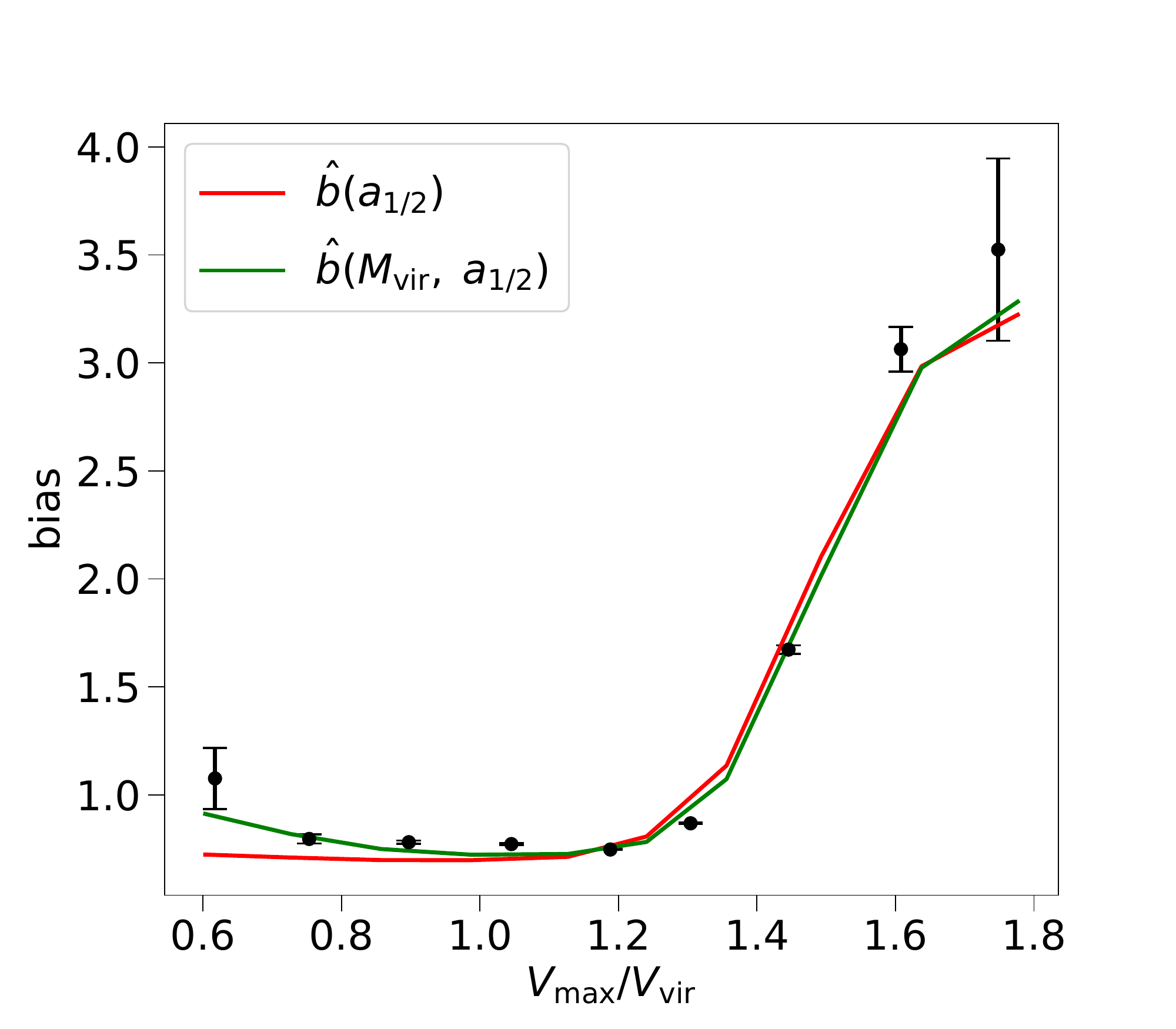}\includegraphics[width=0.4\textwidth]{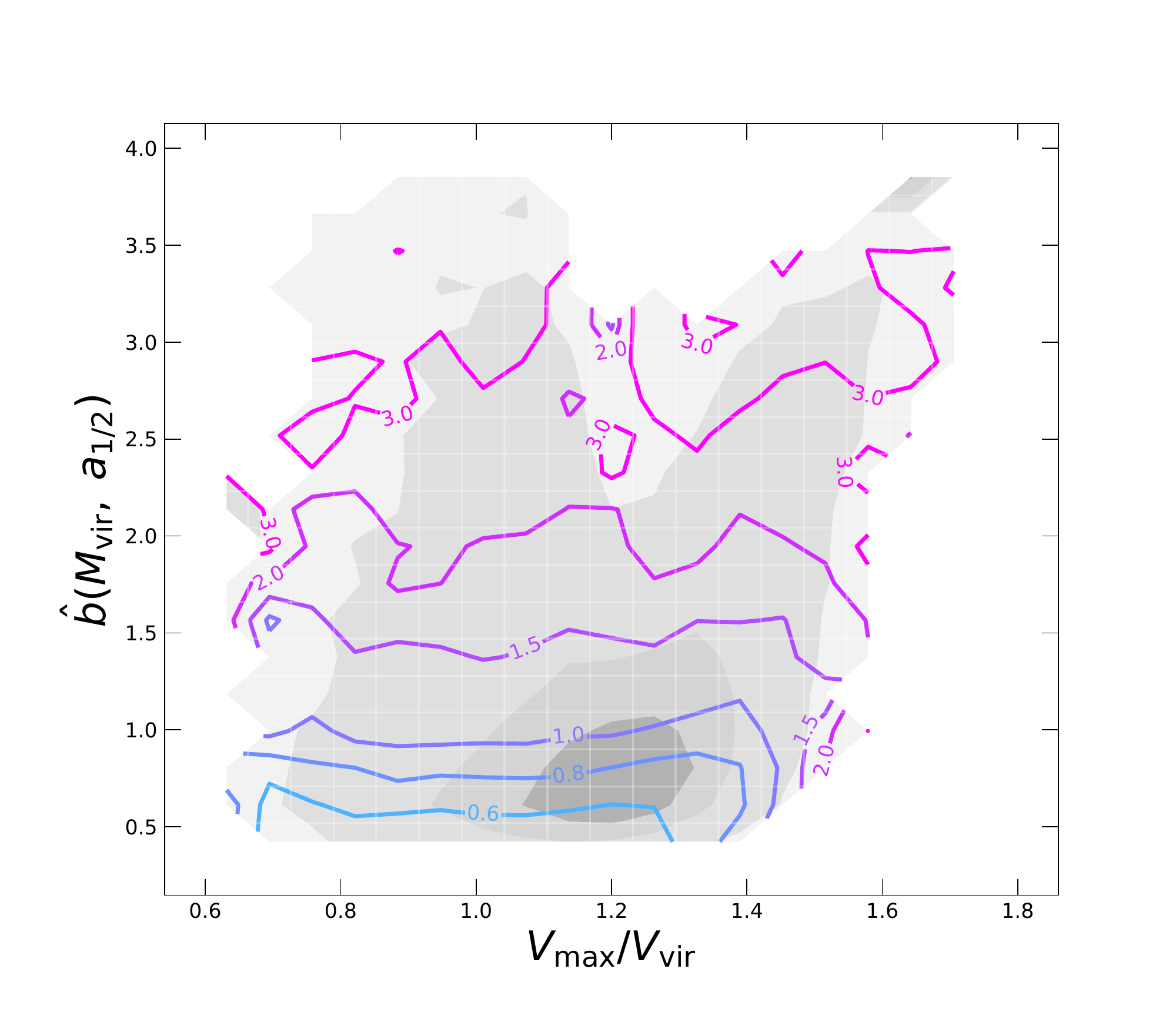}\\
 \includegraphics[width=0.4\textwidth]{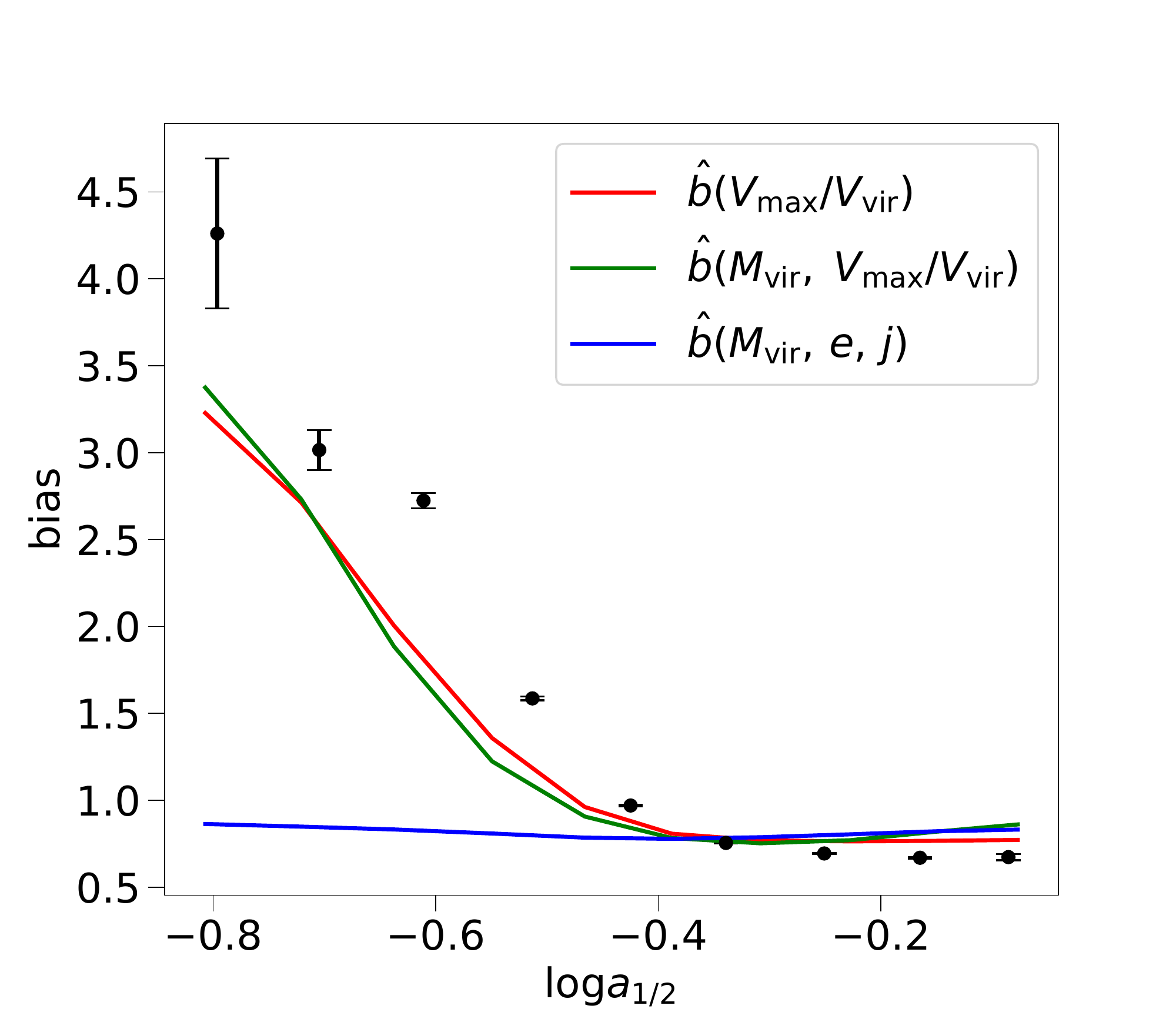}\includegraphics[width=0.4\textwidth]{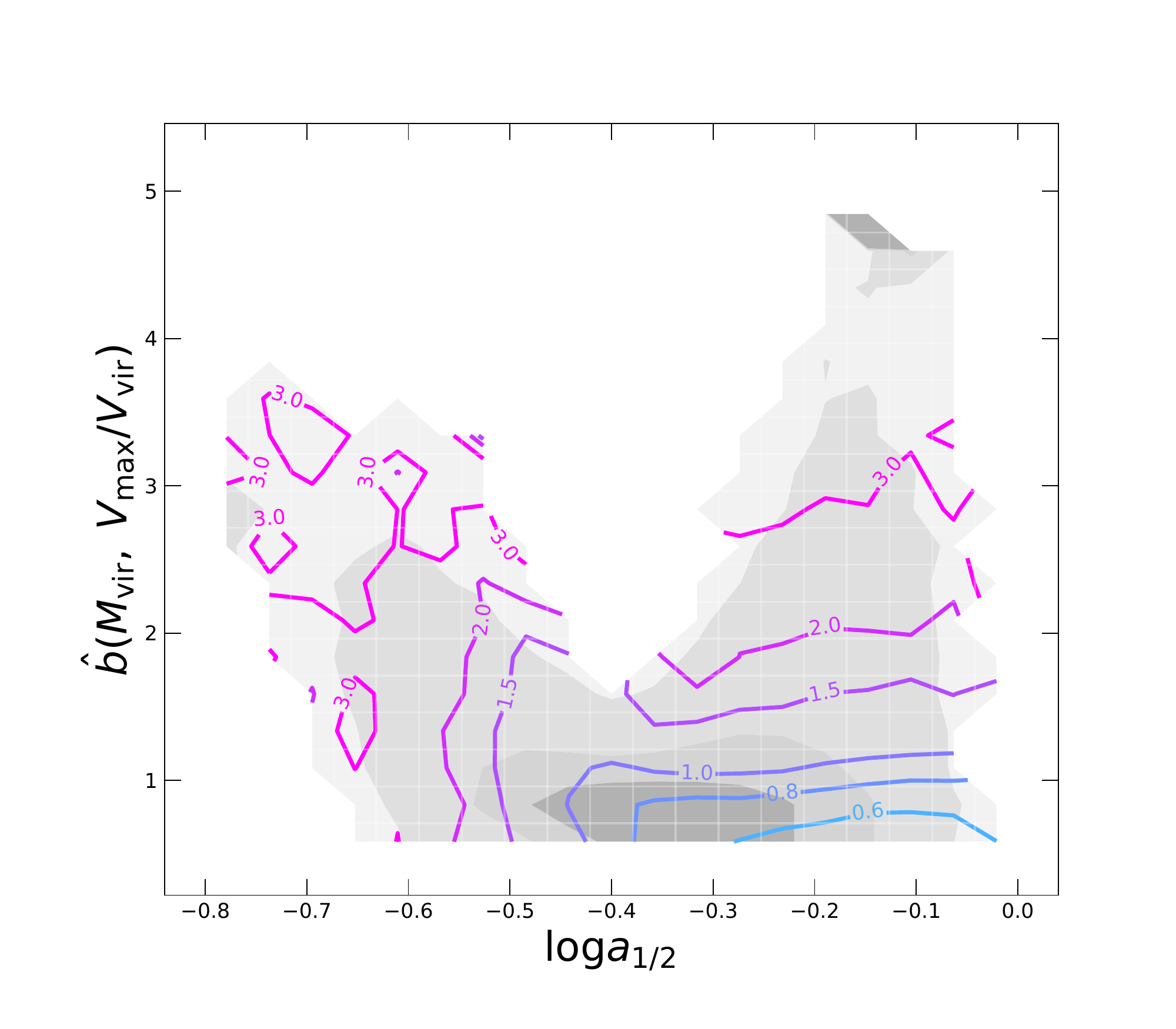}
 \caption{\textit{Top left}: the marginalised dependence of bias on $\vmax$ and the predictions from formation time related estimators. \textit{Top right}: the joint dependence of bias on $\hat{b}(a_\onehalf, M_{\rm vir})$ and $\vmax$. \textit{Bottom left}: the marginalised dependence of bias on formation time and the predictions from $\vmax$ related estimators. \textit{Bottom right}: the joint dependence of bias on $\hat{b}(M_{\rm vir}, \vmax)$ and formation time.}\label{fig:vmax_formation}
\end{figure*}

This is further demonstrated in Fig.~\ref{fig:vmax_formation}. The marginalised bias dependence on $\vmax$ can be well predicted by the estimator $\hat{b}(a_\onehalf)$, except for some disagreement at the low $\vmax$ end. Further introducing a mass dependence besides formation time fixes the disagreement, and predicts the $\vmax$ dependence well over the entire range. As shown in the top right panel of Fig.~\ref{fig:vmax_formation}, once the $\hat{b}(a_\onehalf, M_{\rm vir})$ dependence is accounted for, there is almost no residual dependence on $\vmax$. \rev{The sharp dependence on $\vmax$ at $\vmax\sim 1.5$ for a small population of haloes (vertical contours, also visible in the second panel of Fig.~\ref{fig:bias_mah}) is mostly contributed by a population of tidally stripped haloes, which we will discuss more in section~\ref{sec:eject} below.}

Similarly, as shown in the bottom left panel, the marginalised bias dependence on $a_\onehalf$ can also be largely predicted by the $\hat{b}(\vmax)$ estimator, although at a less accurate level than the transposed case in the top left panel. This time, adding a mass dependence does not help to improve the prediction. As shown in the bottom right panel, the $\hat{b}(M_{\rm vir}, \vmax)$ estimator fails to fully reproduce the formation time dependence for early-forming haloes, leaving contour lines determined primarily by $a_\onehalf$ at the low $a_\onehalf$ end. The loss of reproducibility is also evident by comparing their $\gamma$ values (labelled in Fig.~\ref{fig:internal} and Fig.~\ref{fig:bias_mah}): $0.11$ for $\hat{b}(M_{\rm vir}, \vmax)$ is smaller than $0.14$ for $\hat{b}(M_{\rm vir}, a_\onehalf)$. This means the $\vmax$ dependence of bias is driven by the formation time dependence, with some loss of information when going from $a_\onehalf$ to $\vmax$ in bias modelling. In other words, $\vmax$ is a lossy proxy of $a_\onehalf$ for bias modelling, which is further contaminated by $M_{\rm vir}$.

Such loss of information is not compensated by other internal halo properties either. As shown in the bottom left panel, the three dimensional estimator without $\vmax$ completely fails to predict the formation time dependence at low $a_\onehalf$. This means the mass, spin and shape of haloes barely inherit any information about the formation time dependence of bias. 


\begin{figure}
 \includegraphics[width=0.5\textwidth]{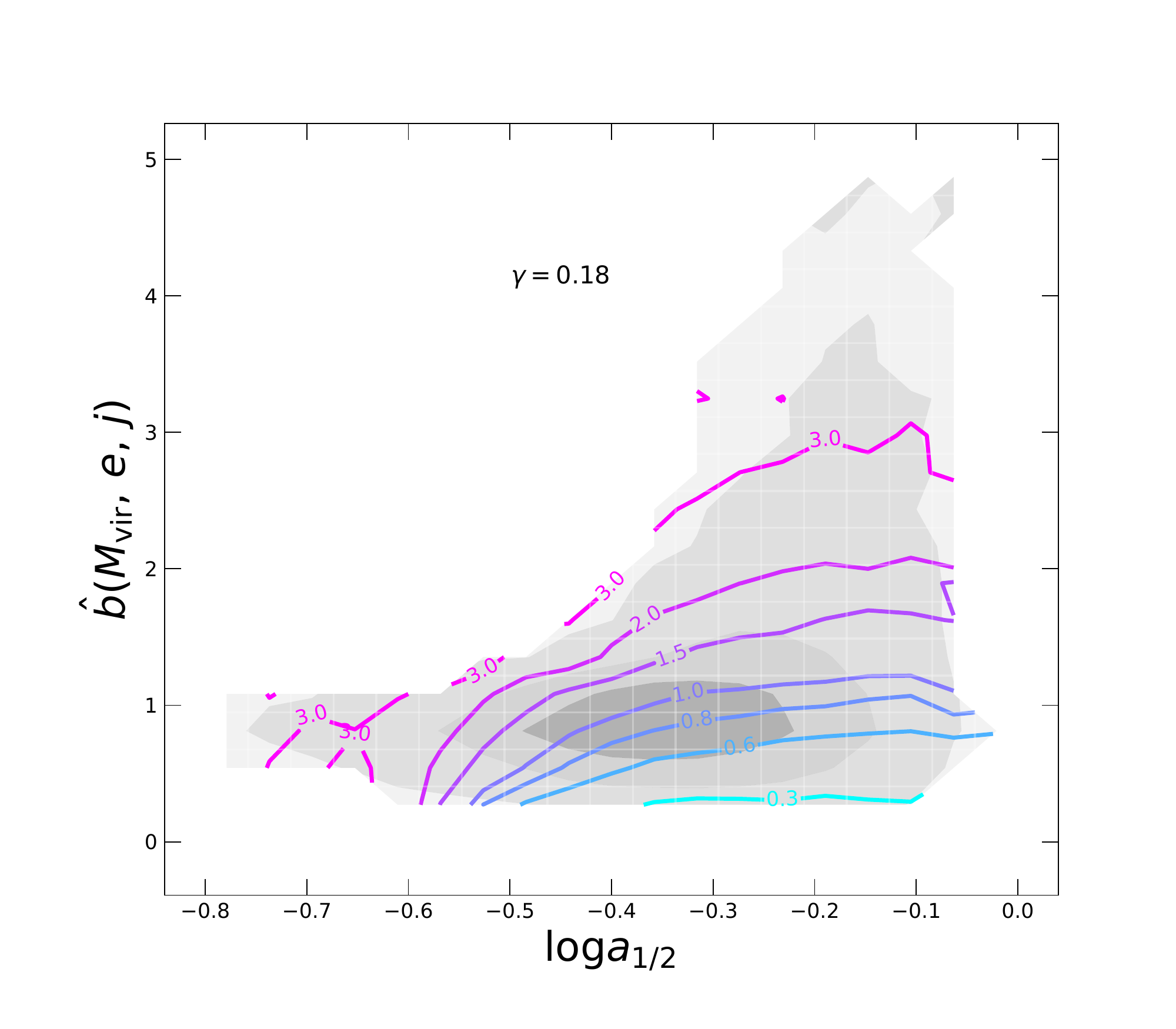}
 \caption{The joint dependence of bias on formation time and a three dimensional bias estimator that models internal halo properties. The $\gamma$ value show the correlation coefficient between $\beta$ and the GPR fit $\hat{b}(M_{\rm vir},\, e,\, j,\, a_\onehalf)$ constructed from this map. 
 }\label{fig:formation_3d}
\end{figure}

Now that $\vmax$ is redundant in presence of $a_\onehalf$ and $M_{\rm vir}$, we can drop it and focus on the bias dependence on the remaining properties. Replacing $\vmax$ with $a_\onehalf$ and combine it with the other three internal properties, we can obtain an estimator that correlates with $\beta$ at a level of $\gamma=0.18$, as shown in Fig.~\ref{fig:formation_3d}. It is also evident that the bias of late-forming haloes barely varies with formation time and only depends on mass, spin and shape. 

Following the procedures in section~\ref{sec:internal}, we have further verified that the formation time, mass, spin and shape are non-redundant in modelling the bias. Our result is consistent with those of \citet{Mao17} who found that the secondary bias in cluster haloes almost does not depend on formation time but more on other parameters. Their findings question the interpretation and naming of these dependences as ``\emph{assembly} bias''. Our result is more general and \rev{demonstrates that the shape and spin dependence of bias is different from the formation time dependence.} This implies that it would be inappropriate to use ``assembly bias'' to describe the spin- and shape-dependent bias, \rev{at least not when the assembly bias is ascribed to the half mass formation time dependence.}


\subsubsection{The minor contribution from ejected haloes}\label{sec:eject}
\begin{figure}
 \includegraphics[width=0.5\textwidth]{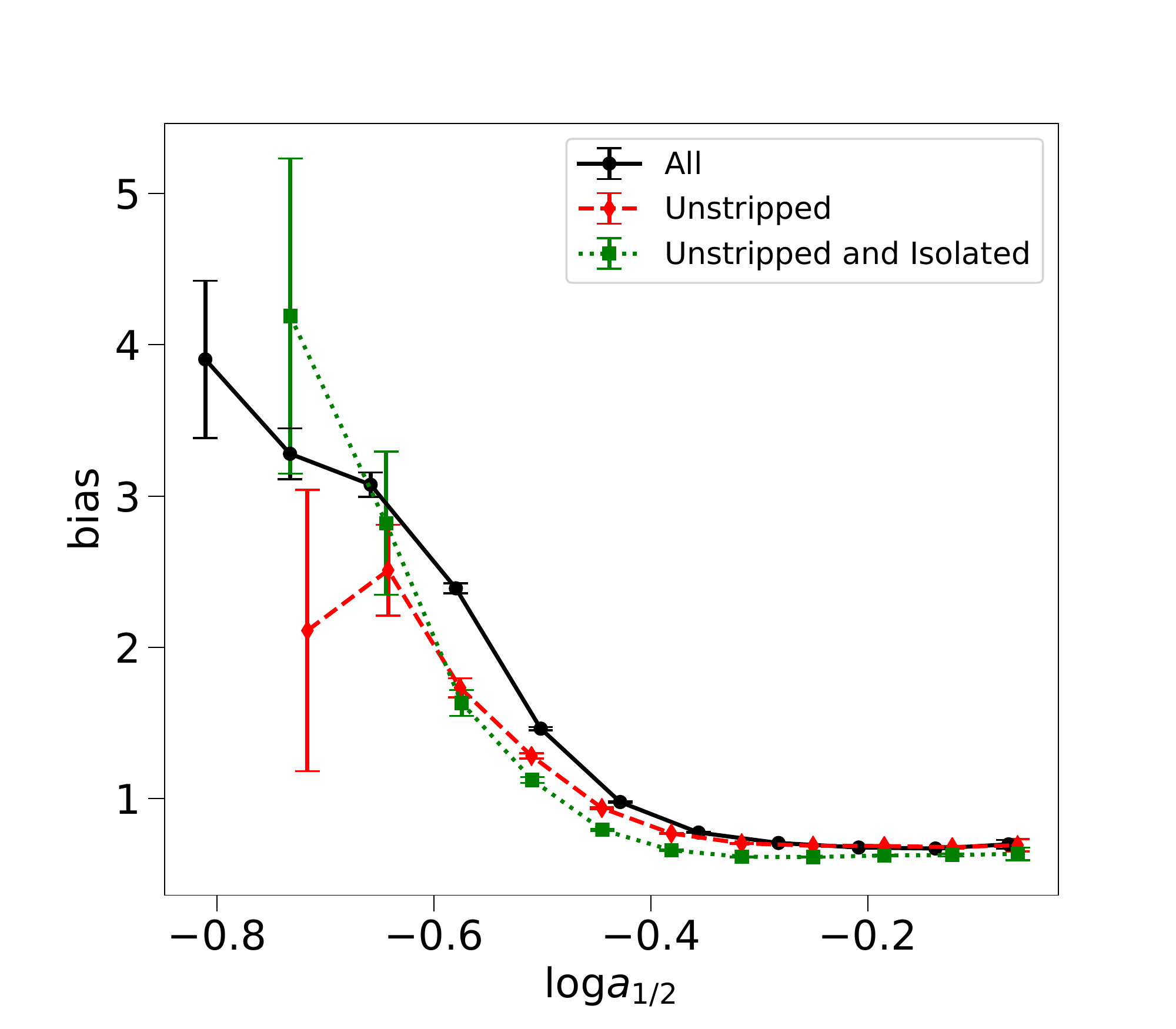}
  \caption{The marginalised bias dependence on formation time. The black curve shows the result without any selection. The `unstripped' population in red dashed line refer to haloes whose peak masses do not exceed their final masses. The green dotted line further removes haloes that are located inside three times the virial radius of a more massive halo.
  }\label{fig:ejection}
\end{figure}

It has been argued that the assembly bias in low mass haloes are driven by a population of haloes ejected from larger ones~\citep{Dalal08}. These haloes are expected to have a steeper outer profile and an earlier half-mass time due to tidal stripping from the previous host halo that could suppress or revert their growth. \citet{Dalal08} argued that these ejected old haloes are expected to follow the motion of the large scale bulk flow and thus reaching an unbiased state. However, we find it inconsistent with our high resolution simulation that resolves a highly biased population of old haloes.

In Fig.~\ref{fig:ejection} we show the influence of this population on the formation time dependence explicitly.
We try two criteria to remove the ejected population, similar to that used in \citet{Dalal08}. First, we keep only unstripped haloes by requiring the maximum mass of a halo along its evolution history to not exceed its final mass. In addition to this, we also try a more strict selection that further removes haloes that are located inside three times the virial radius of a more massive halo. As shown in the figure, such selections eliminates the oldest haloes with $\log a_\onehalf\sim -0.8$. It also tends to slightly lower the bias function, because the removed haloes are in the vicinity of more massive haloes and hence more biased. Despite this small difference, the overall formation time dependence is largely unaffected, in good agreement with \citet{Wang09} who found that the assembly bias of low mass haloes is not mainly determined by ejected haloes. Rather, the formation time dependence is consistent with a more general picture that the earlier forming haloes tend to live in a denser environment~\citep{Wang07} and thus more biased. We will return to the environmental dependence in more detail in section~\ref{sec:env}.

\subsection{Beyond a single formation time dependence}

\begin{figure*}
 \includegraphics[width=\textwidth]{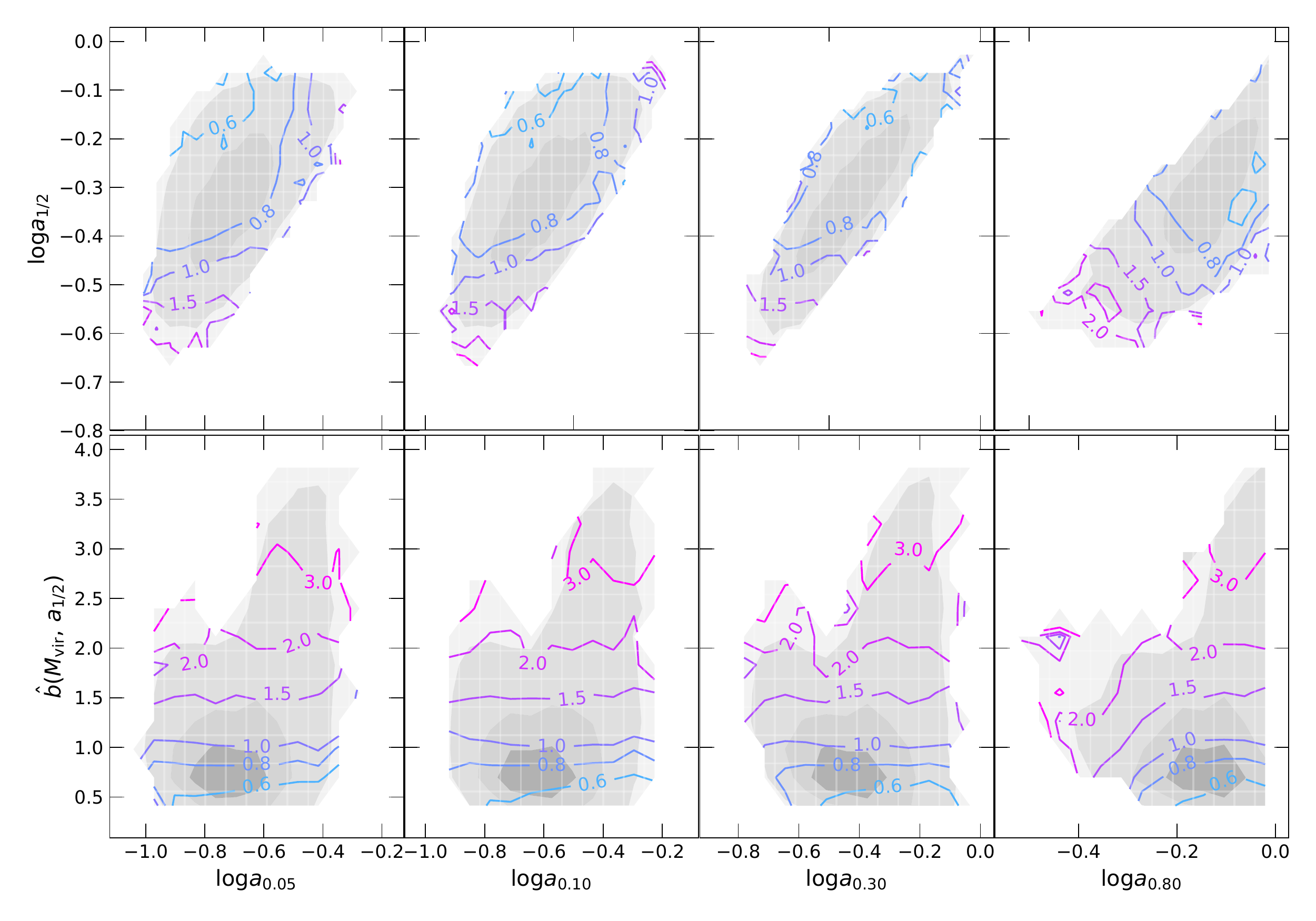}
 \caption{Joint dependence of bias on different formation time measures. Only the `unstripped' haloes whose peak mass does not exceed its final mass are used in this plot.}\label{fig:formation}
\end{figure*}

In Fig.~\ref{fig:formation} we explore the dependence of bias on alternative definitions of halo formation time, defined at different fractional masses. As shown in the top panels, for early-forming haloes, the bias is mostly determined by the recent growth history, while the formation time defined at a much smaller mass fraction than $1/2$ almost does not matter. For late-forming haloes, there appears to be a dependence on the early formation history. However, as is shown in the bottom panels, this dependence disappears once we account for the mass dependence. 
\rev{After accounting for both the mass and $a_\onehalf$ dependence, for early-forming haloes, there is still a dependence on the very recent MAH as revealed by the $a_{0.8}$ parameter. Combined with previous analysis on the $(M_{\rm vir}, a_{\onehalf})$ dependence, these plots show that the bias of late-forming haloes ($\log a_\onehalf>-0.3$) are almost independent of the MAH, while the bias of early-forming haloes are best determined by their recent MAH. Note that haloes with an early $a_{0.8}$ means the mass accretion at a late time is highly suppressed, making the MAH to deviate from a single parameter family. 
} 
We have only used unstripped haloes in this figure. For haloes that experienced mass stripping, the bias also tends to depend on multiple formation time parameters defined at high mass fractions. These objects can be regarded as an extension of the population with suppressed recent MAHs, whose $a_{0.8}$ can be even smaller and whose biases are also higher.

\rev{This analysis however does not completely rule out the possible existence of a MAH-dependent bias (or assembly bias) in late-forming haloes. In particular, the existence of an assembly bias in cluster haloes is well-motivated in theory~\citep[e.g.][]{Dalal08}. Very recently \citet{Chue18} has also tried an optimal transformation of the MAH into an assembly variable to show the existence of such a bias using a much larger sample of cluster haloes. Our analysis of combining two different formation time variables suggest that any MAH dependence in late-forming haloes including clusters must be sufficiently weak and subdominant compared to the dependences on mass, spin and shape. In an upcoming work, we will extend our analysis to also study the Lagrangian properties of haloes to further understand the relation between various bias factors and the formation mechanism of haloes.}

\section{Dependence on non-parametric density profile: the emergence of an environment scale}\label{sec:env}
\begin{figure*}
 \includegraphics[width=\textwidth]{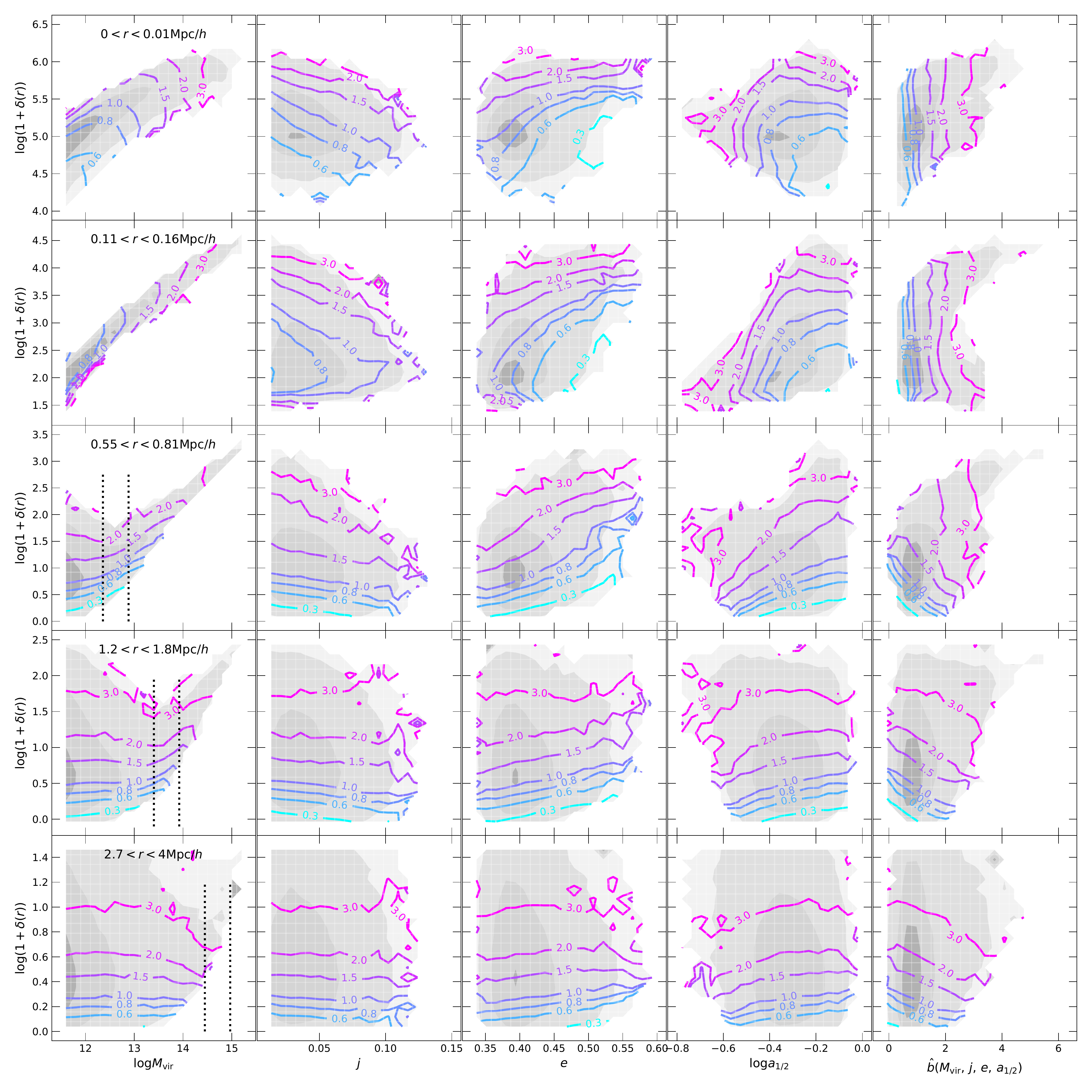}
 \caption{The joint dependence of bias on halo properties and halo density profile $\delta(r)$. Each column shows the dependence on one halo property, while different rows show the dependence on the density profile measured at different radius as labelled. The vertical dotted lines in the first column marks the mass range of haloes whose $2R_{\rm vir}$ falls in the radial range of the density profile measurement. The two lines correspond to the values evaluated at the two edges of the radial bin. This mass scale is located to the left of the axis limit in the first two rows and thus not shown. The last column shows the dependence on the four dimensional estimator constructed in Fig.~\ref{fig:formation_3d}.}\label{fig:halo_env}
\end{figure*}

After accounting for the dependences on the internal parameters, it remains interesting to see if there is still significant information contained in the non-parametric density profile of a halo.
In Fig.~\ref{fig:halo_env} we show the dependence of bias on the density profile $\delta(r)$ and other internal parameters. Given that $\vmax$ is a lossy proxy of $a_\onehalf$ as shown in Sec.~\ref{sec:mah}, we will use $a_\onehalf$ in place of $\vmax$ as a structure parameter. The density profiles are measured in spherical shells at various radii from the halo centre. The vertical dotted lines in the first column show the mass scale of haloes whose boundaries are comparable to the radial scale, $r$, of the density profile estimate: haloes to the left of the dotted lines have $2R_{\rm vir}<r$ while those on the right have $2R_{\rm vir}>r$ (see appendix~\ref{app:boundary} for a further justification of the $2R_{\rm vir}$ choice).

In the first two rows, the densities are measured on very small scales well inside the boundaries of all the haloes in our sample, and there are significant dependences on all the internal parameters after controlling $\delta(r)$. Although we have not modelled the density dependence in the combined estimator $\hat{b}(M_{\rm vir},\, j,\, e,\, a_\onehalf)$, it successfully accounts for the dependence on $\delta(r)$ inside haloes as shown in the top two panels of the last column. This suggests that this estimator captures most, if not all, of the bias information contained in the \rev{internal density field} of haloes.

As the density is controlled on a larger and larger scale, the dependences on all the internal halo properties weaken. In the extreme case when the density is measured at the radial scale $r_b$ where the bias is defined
, the bias $\beta(r_b)=\delta(r_b)/\xi(r_b)$ depends completely on $\delta(r_b)$ by construction, and no secondary dependence on any other parameters will be left. Practically, for $r>1\mpch$, the dependence on other parameters become subdominant after accounting for the density dependence. Note that as the scale becomes larger it reflects more of the large scale clustering itself and is less relevant to properties of haloes. As a result, we have adopted the minimum scale at $1\lesssim r\lesssim 2\mpch$ to define an ``environmental'' density of dark matter haloes.

\begin{figure*}
 \includegraphics[width=\textwidth]{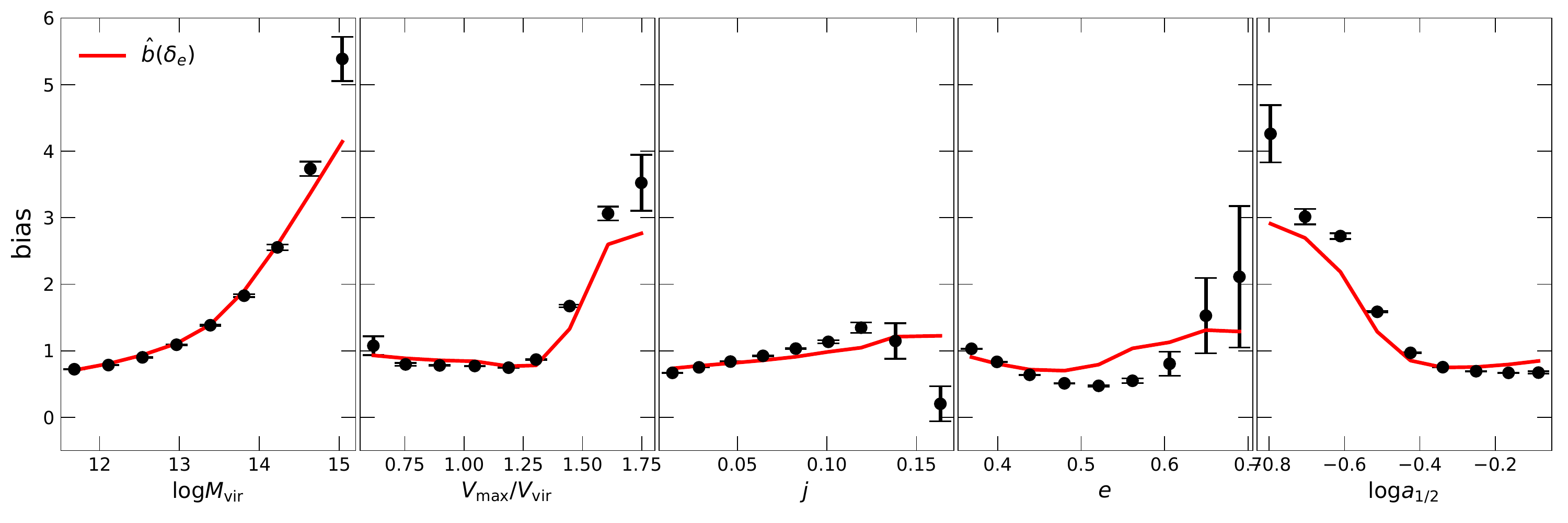}
 \caption{The dependences of bias on internal parameters (data points) and the predictions (solid lines) from a GPR model that only models the environmental dependence of bias.}\label{fig:GPR_env}
\end{figure*}

As the environment accounts for the majority of the bias dependence on the internal structure, we have fitted a bias estimator that only models the environment dependence using GPR, $\hat{b}(\delta_e)$. The performance of this estimator in predicting the bias dependence on the internal parameters are shown in Fig.~\ref{fig:GPR_env}. As expected from Fig.~\ref{fig:halo_env}, the estimator largely reproduces the dependences on all the internal parameters, even though none of these internal parameter dependences have been modelled explicitly. \rev{Using the multi-scale properties of the Gaussian random field from which haloes are formed, \citet{Shi18} have shown analytically that the bias dependence on the halo mass can be largely absorbed into a dependence on an environmental density defined on a much larger scale than the halo scale. Our findings are consistent with the general picture described in \citet{Shi18}, but further demonstrate that the bias dependence on other halo properties besides mass can also be largely absorbed into an environmental dependence. At the same time, the emergence of an absolute environmental scale of $\sim 1-2\mpch$ is a new discovery and require more theoretical understanding, which we discuss further in section~\ref{sec:discussion}.}

\begin{figure}
 \includegraphics[width=.5\textwidth]{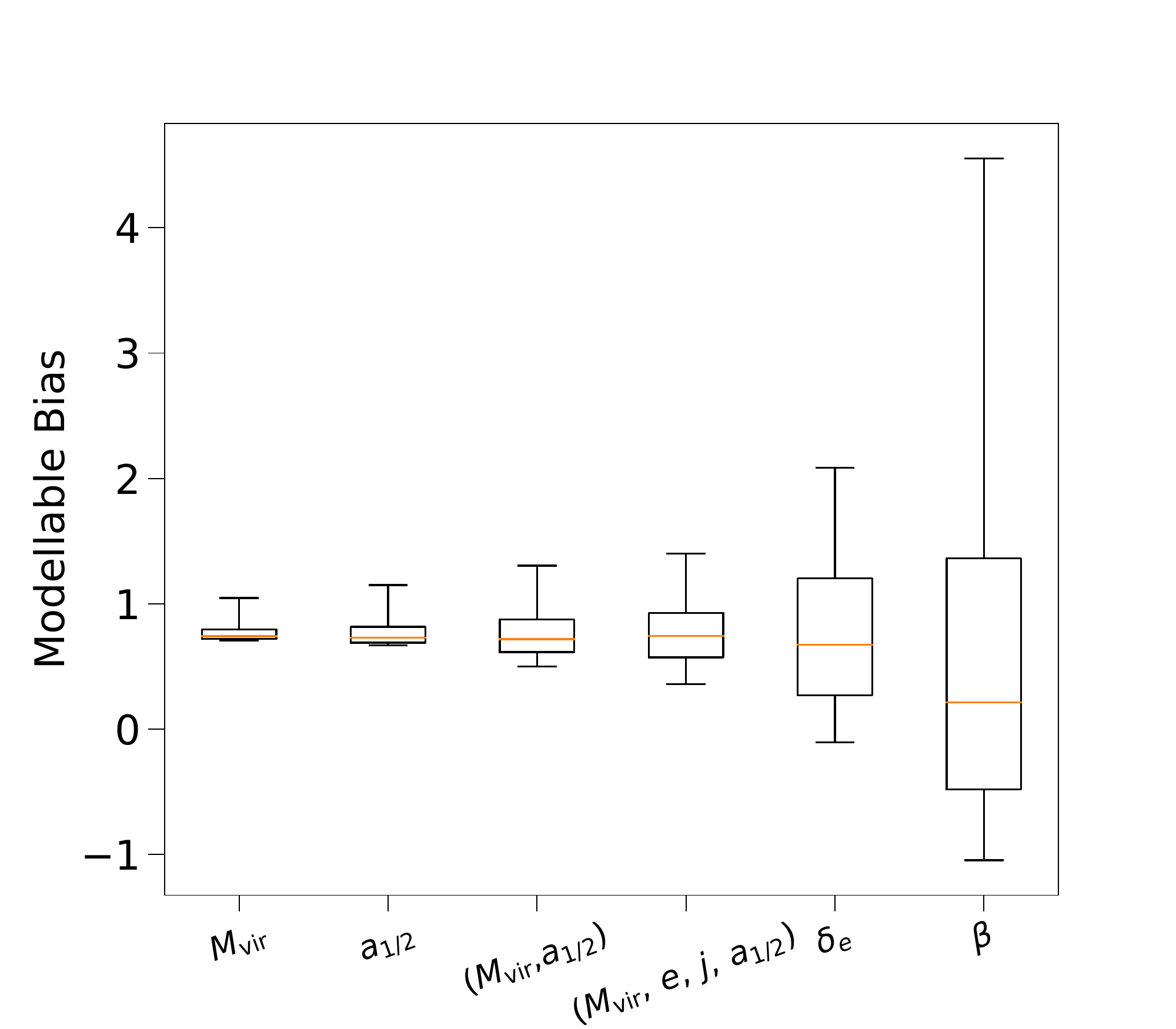}
 \caption{The distribution of predicted (modellable) biases from different bias models. For each bias model, we first predict a bias for each halo in the sample using the corresponding bias estimator. The distribution of the predicted biases from each model is then shown with a box plot, with the bottom and top edges of the box showing the first and third quantiles and the bottom and top ends of the whiskers showing the 5th and 95th percentiles of the distribution. The median value of each distribution is shown as a red bar in the middle of the box. The $x$-axis lists the variables explicitly modelled by each estimator. For reference, the distribution of the individual biases, $\beta$, is shown by the last box.}\label{fig:box}
\end{figure}

We summarise the performance of various bias estimators of different dimensions in Fig.~\ref{fig:box}. For each estimator, the box plot shows the spread in its predicted bias values, which is equivalent to the sensitivity of the estimator in detecting the bias variations across different haloes. When the bias is modelled with a single halo mass parameter, it loses sensitivity in the low mass (low bias) end, leading to very similar predicted biases among low mass haloes. Similarly, the formation time dependence alone is insensitive to biases in the late-forming (low bias) haloes. 
Combining mass and formation time, the bias difference can be better resolved especially among the low bias haloes. Adding constraints from shape and spin further improves the sensitivity of the bias estimator. 
When the bias is modelled from the environment, the sensitivity is the highest. For reference, the distribution of the individual bias of each halo, $\beta$, still has a much larger spread than that from any of the bias models, which means there is still significant amount of fluctuations in $\beta$ that cannot be modelled by the previous estimators.

\rev{Although our bias measurement is made on a relatively small scale of $6-9\mpch$, we emphasise that all the bias dependences explored in this work are not sensitive to the scale of the bias measurement as long as the scale is large enough to capture the linear bias. We have tested that adopting a much larger scale at $\sim 30\mpch$ to measure the bias gives almost identical results in the bias dependences, except that the scatter of $\beta$ becomes larger, reflecting a larger fraction of stochasticity in the large-scale clustering.}

\begin{figure}
 \includegraphics[width=0.5\textwidth]{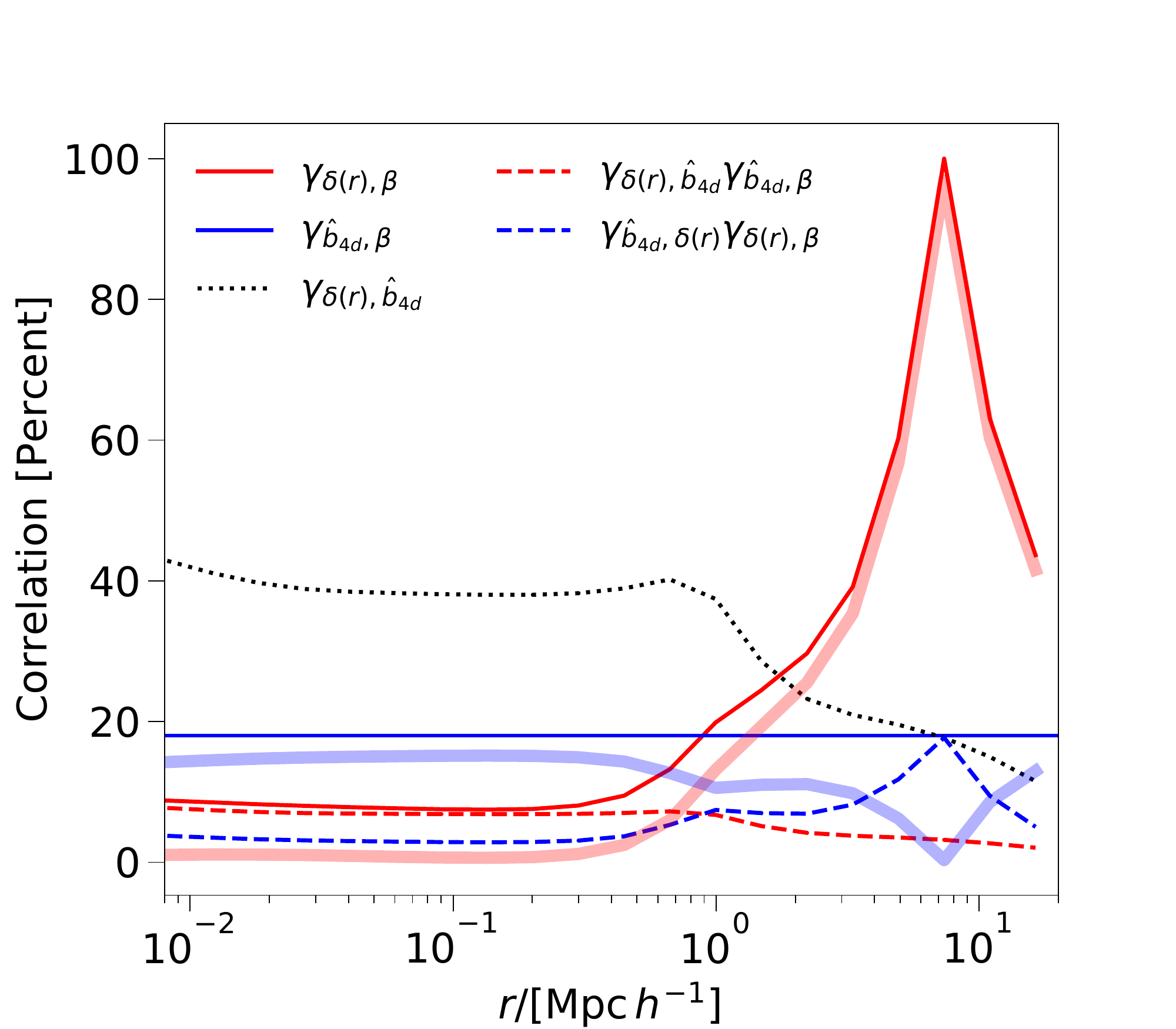}
 \caption{The correlations between the internal estimator, $\hat{b}_{4d}$, density profile, $\delta(r)$, and bias, $\beta$. The internal estimator refers to the four dimensional estimator constructed in Fig.~\ref{fig:formation_3d} that models the full dependence of bias on internal parameters of haloes. The thin solid lines show the full correlations between the corresponding quantities as labelled. The dashed lines show the correlated components, with the red dashed line showing the component of $\gamma_{\delta(r),\beta}$ that can be accounted for through the $\hat{b}_{4d}$ dependence, and the blue dashed line showing the component in $\gamma_{\hat{b}_{4d},\beta}$ that can be accounted for through the density profile dependence. The thick solid lines show the independent components that cannot be explained by the other variable (i.e., the differences between the thin solid and dashed lines). The dotted line shows the correlation between the internal estimator and the density profile.}\label{fig:corr_prof}
\end{figure}

\section{Discussions}
\label{sec:discussion}

\subsection{The relative importance of environment and internal structure}

To assess the relative importance of environment and internal structure in a more quantitative way, we decompose the correlation between $\delta(r)$ and $\beta$ into a component explainable by the internal structure dependence of bias, and a residual component that is independent of the internal estimator. The decomposition is done as follows. Assuming the dependence of bias on $\delta(r)$ is only introduced through its correlation with an estimator $\hat{b}$ (which is $\hat{b}_{4d}$ in this case) that satisfies $\beta=\hat{b}+\epsilon$ where $\epsilon$ is a random variable independent of $\hat{b}$ (i.e., $\delta(r)\rightarrow\hat{b}\rightarrow\beta$ is a Markov process), it immediately follows that
\begin{align}
 \gamma_{\delta(r),\beta}&=\gamma_{\delta(r),\hat{b}}\frac{\sigma_{\hat{b}}}{\sigma_{\beta}}\nonumber\\
 &=\gamma_{\delta(r),\hat{b}}\gamma_{\hat{b},\beta}.\label{eq:corr_propagate}
\end{align}
Thus $\gamma_{\delta(r),\hat{b}}\gamma_{\hat{b},\beta}$ measures the amount of correlation between $\delta(r)$ and $\beta$ propagated through the intermediate variable $\hat{b}$, and the residual $\gamma_{\delta(r),\beta}-\gamma_{\delta(r),\hat{b}}\gamma_{\hat{b},\beta}$ gives the component independent of $\hat{b}$. An analogous decomposition is done for the correlation between the internal estimator and $\beta$ as well.\footnote{Strictly speaking, the decomposition of $\gamma_{\hat{b},\beta}$ should be done using $\hat{b}(\delta(r))$ as an intermediate variable instead of $\delta(r)$. However, we have checked that the two produces nearly identical results, since the correlation with $\delta(r)$ is quite close to the correlation with $\hat{b}(\delta(r))$.} We provide a more general discussion of the mathematics of the correlation coefficient decomposition in Appendix~\ref{app:corr_sum}.

As shown in Fig.~\ref{fig:corr_prof}, on small scales ($r<0.3\mpch$), the bias correlation of the density profile can be almost fully explained by the internal estimator dependence,  $\gamma_{\delta(r),\beta}\approx \gamma_{\delta(r),\hat{b}_{4d}}\gamma_{\hat{b}_{4d},\beta}$. On larger scales, the density profile correlation with bias increases rapidly, while the contribution from the internal estimator decreases due to a decrease in the correlation between internal structure and external density profile, $\gamma_{\delta(r),\hat{b}_{4d}}$.
On the other hand, only a tiny fraction of the internal estimator dependence can be accounted for by the density profile dependence on small scales. The amount of $(\hat{b}_{4d},\beta)$ correlation explainable by $\delta(r)$ increases with scale, because of the steep increase in the similarity between $\delta(r)$ and $\beta$. Overall, the amount of independent bias information contained in the internal estimator and the density profile becomes comparable at around $1\mpch$, and is dominated by the density profile dependence on larger scales. Note that even with an environment defined at $r\sim 3\mpch$, there is still $\sim 10\%$ variation in $\beta$ that can be solely attributed to the internal estimator dependence, although the amount attributable solely to the density profile dependence is much higher. This is consistent with the negligible but detectable internal estimator dependence in the lower right panel of Fig.~\ref{fig:halo_env}.

\subsection{Choice of the environment scale: is it better to use a relative scale?}
\begin{figure}
 \includegraphics[width=0.5\textwidth]{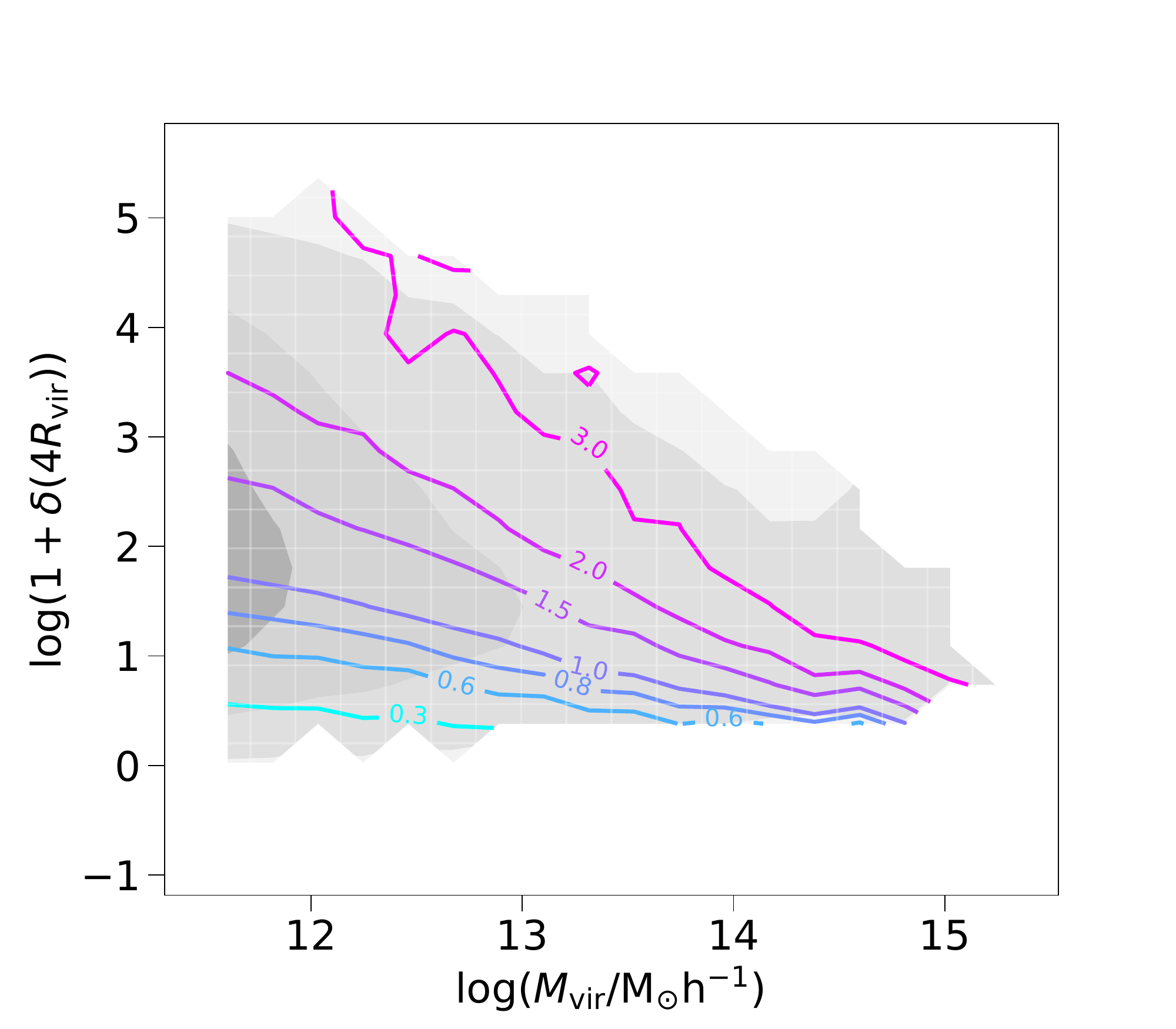}
 \caption{The joint dependence of bias on halo mass and external density at $4R_{\rm vir}$ of each halo.}\label{fig:halo_env_rel}
\end{figure}

\rev{According to the first column of Fig.~\ref{fig:halo_env}, the dependence on mass at a fixed $\delta(r)$ is removed for all the haloes that satisfy $2R_{\rm vir}<r$, while the mass dependence for larger haloes remain. It is tempting to test whether we can further remove this residual dependence by defining the environmental density at a radius relative to the size of each halo. The answer is negative according to Fig.~\ref{fig:halo_env_rel}, which shows that the bias depends significantly on both mass and environment if the environment is defined at $4R_{\rm vir}$ of each halo. For haloes with the same bias, the low-mass ones are surrounded by higher densities at $4R_{\rm vir}$, while the high-mass ones are surrounded by lower densities at $4R_{\rm vir}$. Defining the environment at other relative distances outside haloes produces similar results. This is not difficult to understand from the results of Fig.~\ref{fig:halo_env}. As shown by the nearly horizontal contour lines in the last two rows of Fig.~\ref{fig:halo_env}, haloes with similar biases will have similar large-scale density profiles that also decrease with radius. A high-mass halo has a larger $R_{\rm vir}$ and hence its environment density $\delta(4R_{\rm vir})$ is measured on a larger absolute scale, which naturally leads to a lower density. }

\begin{figure*}
 \includegraphics[width=0.8\textwidth]{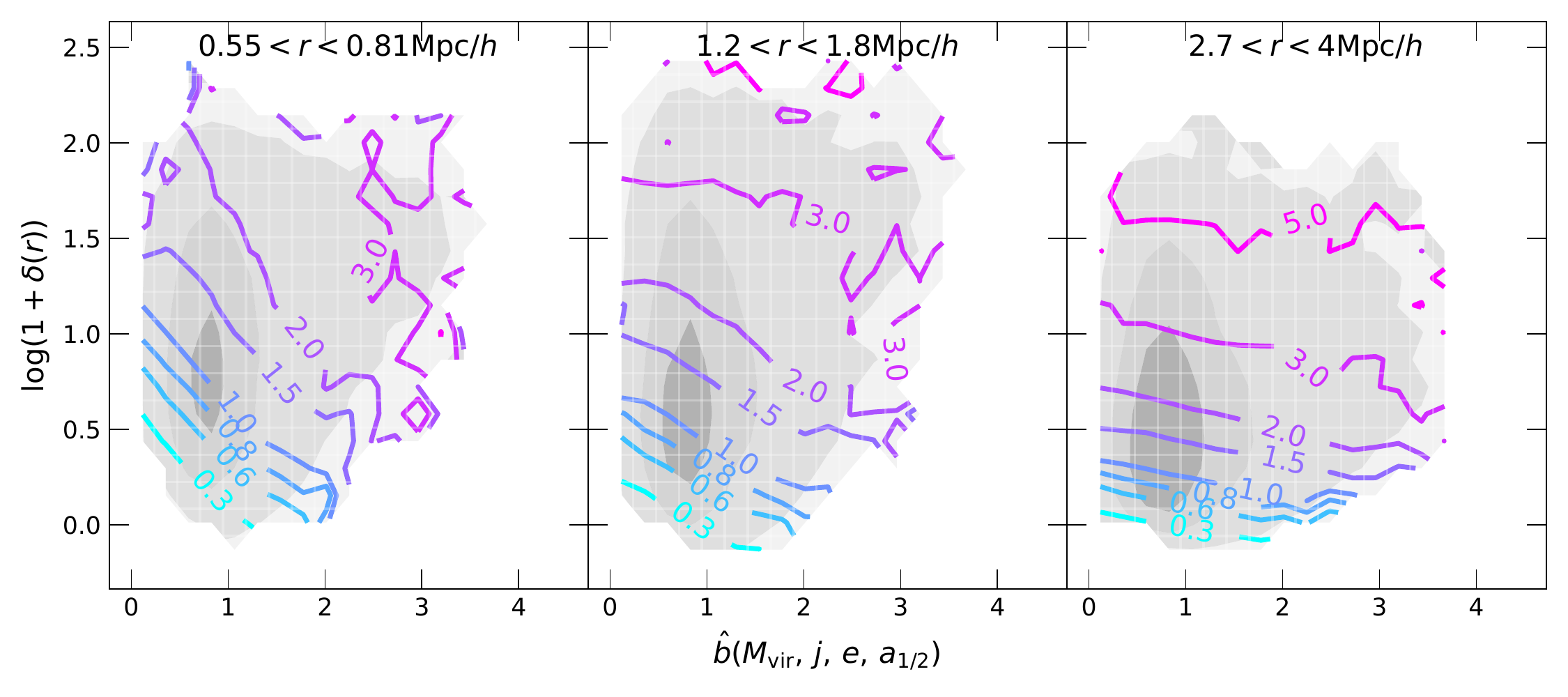}
 \caption{The joint dependence of bias on halo properties and halo density profile $\delta(r)$. The three panels measure $\delta(r)$ at three different radial scales. Only haloes that satisfy $r>2R_{\rm vir}$ (i.e., haloes to the left of the first vertical lines in the corresponding panels of Fig.~\ref{fig:halo_env}) are used. The horizontal axis is a bias estimator that explicitly models the bias dependence on halo mass, spin, shape and formation time.}\label{fig:halo_env_split}
\end{figure*}

\revthree{Even though the attempt to define an environment at a relative radial scale fails, it still remains interesting to test whether we can define the environment at a smaller scale if we only study a sample of smaller haloes. This works in the case of the mass dependence according to Fig.~\ref{fig:halo_env}: for a sample of low mass haloes, we can define the environment, $\delta(r)$, at a smaller $r$ as long as $2R_{\rm vir}<r$ is satisfied. When this $\delta(r)$ is controlled the bias no longer depends on mass. In Fig.~\ref{fig:halo_env_split} we test whether this environemtal density also works to remove the bias dependence on other halo properties. We show the joint dependence on $\delta(r)$ and the four-dimensional estimator, for haloes that are located to the left of the first vertical line in the corresponding panels of Fig.~\ref{fig:halo_env}. Same as in Fig.~\ref{fig:halo_env}, when the density is controlled at a smaller scale, the dependence on internal parameters is still significant, even though the density is now measured outside $2R_{\rm vir}$ for all the haloes in Fig.~\ref{fig:halo_env_split}.}

These exercises show that it is the absolute radial scale that determines the residual bias dependence on the internal structure. In the language of a halo model~\citep{HaloModel}, the large-scale density distribution around a halo is dominated by the two-halo term, and our minimum radial scale of $\sim 1-2\mpch$ can be understood as the minimum scale beyond which the two halo term is largely independent of the one halo term \revthree{and mostly determined by the environmental density. The dependence of clustering on the halo properties then enters through the correlation between halo properties and the environment.} We will present more detailed studies of the scale-dependence of the biases around and below this scale in a forthcoming paper.

\section{Summary and conclusions}~\label{sec:summary}
The bias of dark matter haloes depends on many halo parameters. In this work, we aim at understanding the interplay of the various bias factors through a comprehensive analysis of the joint dependence of bias on multiple halo parameters. 
Our method starts from a microscopic bias defined for each halo according to its large scale density profile, and subsequently produces bias maps in halo property space by ensemble averages. We adopt a flexible machine learning technique, Gaussian Process Regression (GPR), to smoothly interpolate the bias maps and construct multivariate bias estimators. 
The individual bias estimate and the derived bias estimators also allow us to quantify the sensitivity of bias to various halo properties and their combinations via a correlation analysis.

We apply our method to a dark matter only simulation to study the bias dependence on three families of halo properties, including the internal structure, formation history and environmental densities. The structure parameters include the halo mass, halo shape, halo spin, and $\vmax$ quantifying the shape of the density profile. For formation history, we have studied the half-mass time as well as the formation time defined at different stages of halo evolution. Lastly, we have studied the non-parametric density profile of the haloes from small to large scales. In particular, we focus on the external density defined near the boundary of haloes as an environmental density.

Focusing on a single parameter dependence, we show that although halo mass probes a relatively large dynamic range in halo bias, it is not the most sensitive bias predictor for the majority of haloes due to the weak bias dependence at the low mass end. Among all the properties studied, spin shows the weakest correlation with bias, while environmental density shows the highest sensitivity. The halo shape parameter has a consistently high sensitivity across different mass ranges.

Combining halo mass and another structure parameter, the bias depends mostly on mass for massive haloes, while the dependences on other structure parameters become significant for low mass haloes. Combining all the structure parameters, we show that they are non-redundant with respect to each other in bias modelling. Combining any three of the structure parameters alone cannot reproduce the bias dependence on the remaining fourth parameter.

We explicitly show that the bias dependence on $\vmax$ is driven by formation time dependence. For the first time, we demonstrate that the $\vmax$ dependence can be well explained by the joint dependence on mass and formation time. This is not possible without our new technique that maps out the joint dependences explicitly. On the other hand, we find that the mass, spin and shape of haloes barely contain any information about the formation time dependence of bias. This suggests that it would be inappropriate to use `assembly bias' to describe the spin and shape dependent biases, unless more explicit connections between these biases and the halo formation history are found. In particular, the bias of late-forming haloes is mostly independent of formation time after accounting for the joint dependence on mass, spin and shape, which means the formation time dependence is less important than the other parameters for modelling the bias of late-forming haloes. \rev{For early-forming haloes, the bias depends sensitively on the detailed recent MAH, and a single formation-time parameter defined at the half-mass time does not fully capture this dependence.}


Combining all the internal structure parameters including formation time, $\sim 20\%$ of variation in individual bias on the scale of $6-9\mpch$ can be modelled from the information contained inside haloes. We find that the density profile inside haloes do not provide additional bias information, once we have accounted for the bias dependence on all the internal parameters.

There is an additional large amount of bias information that is not captured by the internal structure, but by the external density field around haloes. Modelling the bias with the environment of haloes defined as the matter density at a halo-centric distance around $1\sim 2\mpch$ greatly increases the predictability of bias, leading to $\sim 30\%$ of predictable bias variation. Such an environment based bias estimator largely reproduces the bias dependences on each of the internal halo parameters, although it does not fully account for the joint dependence on all the internal parameters. The density measured on a larger scale correlates more with the bias, at the expense that it reflects less of a halo property but more of the large scale structure itself. Our environment scale of $\sim 1-2\mpch$ can be understood as the minimum scale beyond which the two halo term of the halo-matter correlation function is largely independent of the one halo term.


Our result has important implications for Halo Occupation Distribution (HOD; \citealp[e.g.,][]{Berlind&Weinberg}) modelling of galaxy clustering. The multivariate bias dependence on internal parameters means that populating galaxies into haloes with different properties other than mass could also make a difference in the clustering signal. 
Thus these nonredundant and bias-sensitive halo properties provide new proxies that can be linked to clustering-dependent galaxy properties. To make this possible, it is desirable to extend our analysis to observed galaxy distribution first to obtain how the bias of galaxies depends on multiple galaxy properties. On the other hand, the prevailing dependence on environment provides a simpler alternative approach for HOD modelling. One may simply focus on populating galaxies into the right environment, to get galaxies with the correct bias. We will explore such possibilities in future works.

\section*{Acknowledgments}
We thank Peter Behroozi, Houjun Mo, Tomomi Sunayama, Uros Seljak and Emanuele Castorina for helpful discussions, and Masahiro Takada and Surhud More for carefully reading the paper and providing detailed comments. This work was supported by JSPS Grant-in-Aid for Scientific Research JP17K14271. YJ is supported by NSFC (11320101002, 11533006, \& 11621303) and 973 Program No. 2015CB857003. TN acknowledges financial support from Japan Society for the Promotion of Science (JSPS) KAKENHI Grant Number 17K14273 and Japan Science and Technology Agency (JST) CREST Grant Number JPMJCR1414.
Kavli IPMU was established by World Premier International Research Center Initiative (WPI), MEXT, Japan.

The authors contribute to this work in the following way. JH and YL initiated the project. JH led the method development, conducted the analysis, and wrote the paper with input from all others. YJ ran the simulation. All authors contributed to discussions and proof-reading the paper.

\bibliographystyle{\mybibstyle}
\setlength{\bibhang}{2.0em}
\setlength\labelwidth{0.0em}
\bibliography{ref}

\appendix
\section{Gaussian Process Regression of bias maps}\label{app:GPR}
The observed bias value $b$ of haloes with a given halo property $x$ (which can be a vector itself, e.g., $x=(M_{\rm vir}, V_{\rm max}, e, j, ...)$) is assumed to follow a normal distribution
\begin{equation}
 b(x) \sim \mathcal{N}\left(g( x), \sigma^2\right),
\end{equation} where $\sigma$ is the measurement uncertainty of $b$ and $g(x)$ is the underlying true bias function. In GPR, $g(x)$ is assumed to be a latent variable following a GP with a covariance function $K(x, x')$ (also called kernel function),
\begin{equation}
 g(x)\sim \mathcal{GP}\left(0, K(x, x')\right),
\end{equation}
such that the joint distribution of any $\left( g(x), g(x^\prime) \right)$ pair follows a two dimensional Gaussian distribution with a zero mean and a covariance of $K(x, x')$. In general, a non-zero mean function can also be adopted, while a zero mean GP is sufficient for our analysis. For the covariance function, we use a Mat\'ern kernel with $\nu=1.5$, which reduces to the following simple form with two free parameters $A$ and $l$,
\begin{equation}
 K(x, x')=A(1+\frac{\sqrt{3}d}{l})\exp(-\frac{\sqrt{3}d}{l}),
\end{equation}
where $d$ is the distance between $x$ and $x'$. We define the distance as the Euclidean distance between the normalised variables $\tilde{x}=x/\sigma_x$ and $\tilde{x}'=x'/\sigma_{x'}$. This is particularly relevant when $x$ represents multiple halo properties, in which case the normalisation is applied separately to each dimension. Such a kernel leads to first-order differentiable Gaussian random fields.

Following this statistical model, the likelihood of the dataset marginalised over $g(x)$ can be written done as a function of the hyper parameters $A$ and $l$. The best-fit parameters are derived maximising the likelihood.

Once the hyper parameters are fixed, the posterior distribution of $g(x)$ can be found to have an expectation of~\citep{GPR}
\begin{equation}
 \mathbb{E}[g(x')|\mathbf{b}(\mathbf{x})]=\mathbf{K}(x', \mathbf{x})[\mathbf{K}(\mathbf{x}, \mathbf{x})+\mathbf{\Sigma}]^{-1} \mathbf{b},
\end{equation} where $\mathbf{b}(\mathbf{x})=(b(x_1), b(x_2), b(x_3),...)$ is the vector of observed biases at a set of halo properties $\mathbf{x}=(x_1, x_2, x_3, ...)$, $\mathbf{\Sigma}$ is the diagonal covariance matrix of the $\mathbf{b}$ with nonzero components $\mathbf{\Sigma}_{ii}=\sigma^2_i$, and $\mathbf{K}(\mathbf{x}', \mathbf{x})$ is the covariance matrix between $\mathbf{g}(\mathbf{x}')$ and $\mathbf{g}(\mathbf{x})$, with $\mathbf{K}_{ij}=K({x'_i}, {x_j})$. This expectation function, $\mathbb{E}[g(x')|\mathbf{b}(\mathbf{x})]$, is the one we use to interpolate the bias maps.

\section{The residual dependences of bias after modelling two internal variables}\label{app:3d}
\begin{figure*}
 \includegraphics[width=0.9\textwidth]{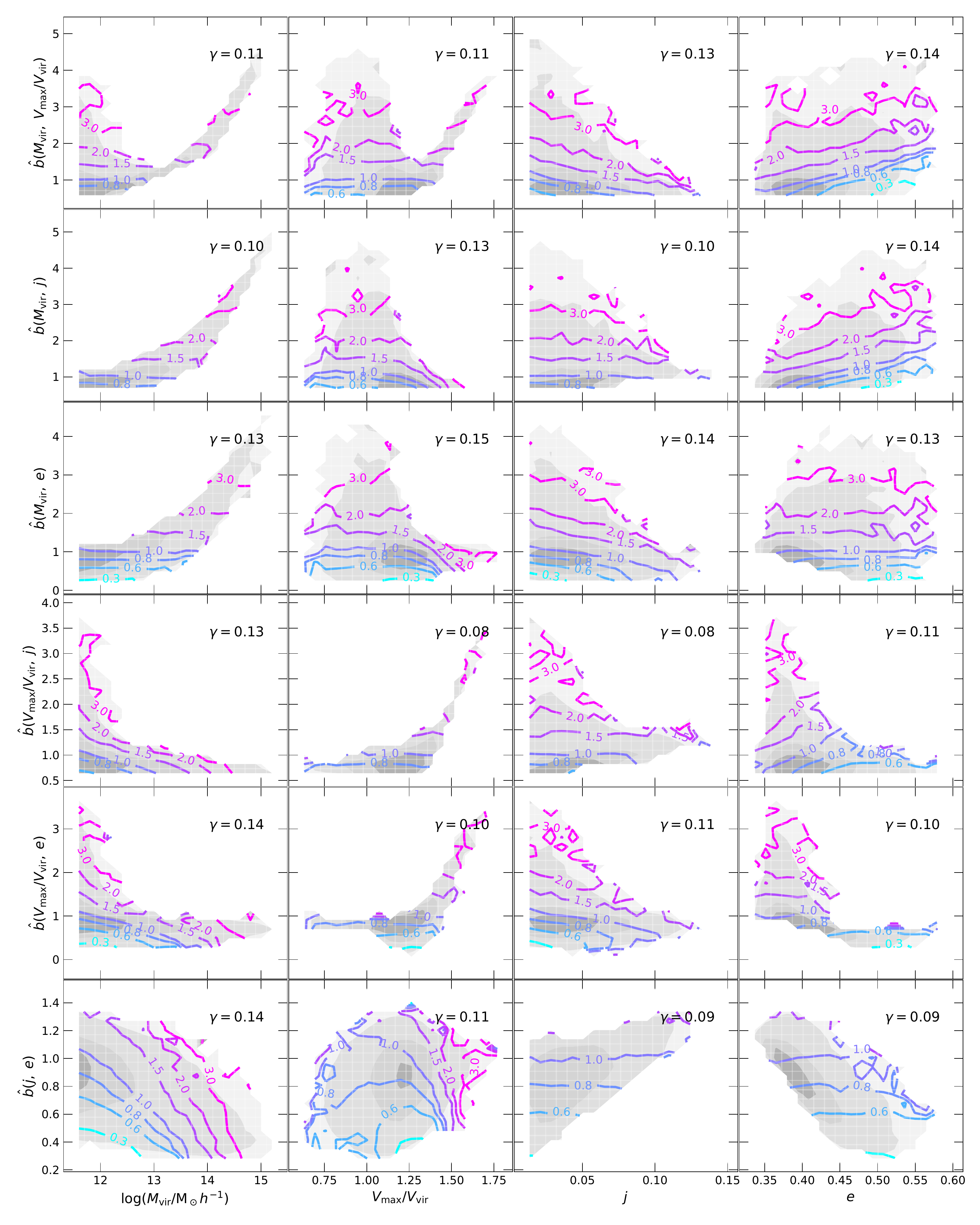}
 \caption{The joint dependence of bias on bivariate estimators and one halo internal property. The $\gamma$ value in each panel shows the correlation coefficient between the GPR estimators $\hat{b}(\hat{b}(x,y),z)$ and $\beta$, where $x$,$y$,$z$ are the halo properties used in the corresponding panel. For clarity, we have not shown the GPR contours.}\label{fig:bias_3d}
\end{figure*}

Fig.~\ref{fig:bias_3d} shows the joint dependences of bias on each bivariate estimator and another structure variable. This time we treat each estimator $\hat{b}(x,y)$ as a derived halo property from $x$ and $y$, and show the bias map of it together with a third halo property $z$. When $z$ is one of the variables involved in the bivariate estimator, $z \in \{x, y\}$, the bias contours are parallel to the $z$ direction, meaning that bias is independent of $z$ for fixed $\hat{b}(x,y)$. This is expected by construction, and again serves to confirm that the GPR estimator $\hat{b}(x,y)$ has fully accounted for the dependence on $z$ if $z \in \{x, y\}$. In contrast, for $z \notin \{x, y\}$ there are always remaining dependences, which appear weak and simple in most cases. The most complex residual dependences are observed in the $(\hat{b}(j,e), M_{\rm vir})$ and $(\hat{b}(j,e), \vmax)$ panels. This can be understood from the overall behaviour of the multivariate bias function: the bias is mostly determined by mass at the high mass end and by formation time (as characterised by $\vmax$, see section~\ref{sec:mah}) for early forming haloes; for low mass and late-forming haloes, bias depends more on $(j,e)$.

For each map, we have constructed a GPR estimator of the form $\hat{b}(\hat{b}(x,y),z)$. The correlation coefficient between each estimator and $\beta$ is also shown in the corresponding panel. Its increment compared with $\gamma_{\hat{b}(x,y), \beta}$ tells the additional bias variation explained by $z$. Note that the correlation for estimators of the form $\hat{b}(\hat{b}(x,y),x)$ or $\hat{b}(\hat{b}(x,y),y)$ are the same as that for $\hat{b}(x,y)$, again confirming that $\hat{b}(x,y)$ has fully described the dependence on $x$ and $y$.

\section{The correlation boundary of haloes}\label{app:boundary}
\begin{figure}
\includegraphics[width=0.5\textwidth]{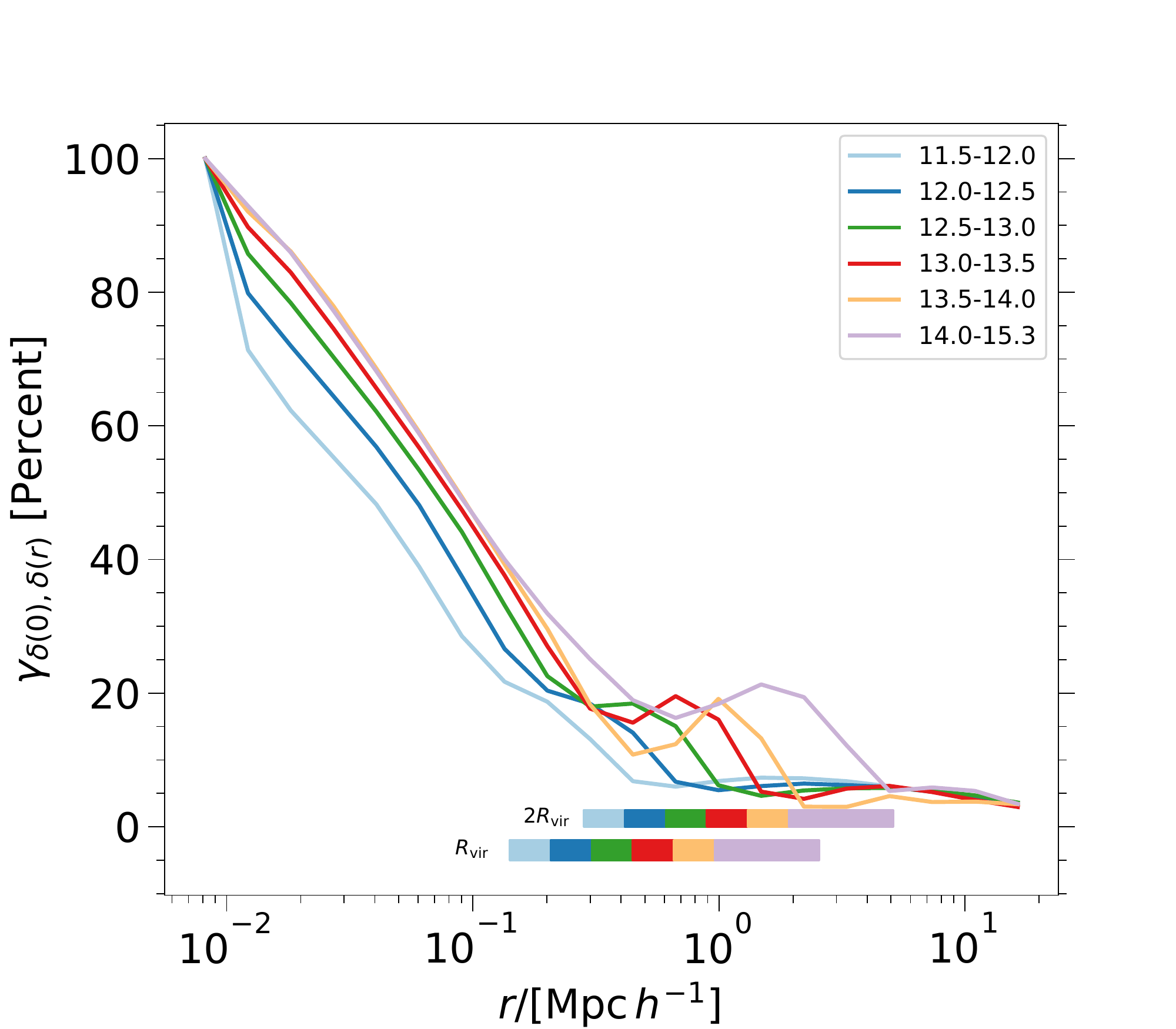}
\caption{The correlation between the central density of haloes and the density at various other radii, for haloes in different mass ranges. The coloured boxes mark the range of $R_{\rm vir}$ and $2R_{\rm vir}$ in each halo mass bin.}\label{fig:boundary}
\end{figure}

In Fig.~\ref{fig:boundary} we show the correlation between the central density of haloes and the density profile. The correlation drops rapidly near the virial radius of the halo, and approaches an asymptotic value of around $5\%$ outside $2R_{\rm vir}$. The behaviour is very similar if we use the density at another small scale radius instead of the central density, except that the location of the peak correlation shifts according to the inner density used.

\section{The correlation coefficient of combined estimators}\label{app:corr_sum}
Suppose $\theta$ and $\eta$ are two independent estimators of $\beta$, such that a combined estimator can be written as $\hat{b}=\theta+\eta$. This combined estimator has a sensitivity
\begin{align}
 \gamma^{}_{\hat{b}\beta}&=\frac{\gamma^{}_{\theta\beta}\sigma_\theta\sigma_\beta+\gamma^{}_{\eta\beta}\sigma_\eta\sigma_\beta}{\sigma_{\hat b}\sigma_\beta}\label{eq:corr_composite}\\
 &=\frac{\gamma^2_{\theta\beta}+\gamma^2_{\eta\beta}}{\gamma_{\hat{b}\beta}}
\end{align} where we have used $\gamma^{}_{\theta\beta}=\sigma_\theta/\sigma_\beta$, $\gamma^{}_{\eta\beta}=\sigma_\eta/\sigma_\beta$ and $\gamma^{}_{\hat{b}\beta}=\sigma_{\hat b}/\sigma_\beta$ in the last equality. This leads to the quadrature addition rule
\begin{equation}
 \gamma^2_{\hat{b}\beta}=\gamma^2_{\theta\beta}+\gamma^2_{\eta\beta}.
\end{equation}

Alternatively, if we substitute $\gamma^{}_{\hat{b}\theta}=\sigma_{\theta}/\sigma_{\hat b}$ and $\gamma^{}_{\hat{b}\eta}=\sigma_{\eta}/\sigma_{\hat b}$ into Equation~\ref{eq:corr_composite}, we obtain the chain rule,
\begin{equation}
 \gamma^{}_{\hat{b}\beta}=\gamma^{}_{\hat{b}\theta}\gamma^{}_{\theta\beta}+\gamma^{}_{\hat{b}\eta}\gamma^{}_{\eta\beta},
\end{equation}
where the two terms are the contributions through the two components of $\beta$. This is a generalisation of Equation~\ref{eq:corr_propagate} in the case that $\beta=\theta+\eta+\epsilon$ with independent $\theta$,$\eta$ and $\epsilon$.
\bsp	
\label{lastpage}
\end{document}